\documentstyle[aps,preprint,prb,eqsecnum]{revtex}

\draft

\newcommand{\bc}{\begin{center}}
\newcommand{\ec}{\end{center}}
\newcommand{\nin}{\noindent}
\newcommand{\be}{\begin{equation}}
\newcommand{\ee}{\end{equation}}
\newcommand{\ba}{\begin{array}}
\newcommand{\ea}{\end{array}}

\newcommand{\dosc}{{d^{\rm osc}}}

\newcommand{\Nosc}{{N^{\rm osc}}}

\newcommand{\Oosc}{{\Omega^{\rm osc}}}

\newcommand{\cl}{\chi_{\scriptscriptstyle L}}
\newcommand{\cgc}{\chi^{\rm \scriptscriptstyle GC}}
\newcommand{\chit}{\chi^{\rm \scriptscriptstyle (t)}}

\newcommand{\kf}{k_{\scriptscriptstyle F}}
\newcommand{\kb}{k_{\scriptscriptstyle B}}
\newcommand{\vf}{v_{\scriptscriptstyle F}}
\newcommand{\bsM}{{\bf \scriptscriptstyle M}}
\newcommand{\lf}{\lambda_{\scriptscriptstyle F}}
\newcommand{\lt}{l_{\scriptscriptstyle T}}

\newcommand{\gs}{{\sf g_s}}

\newcommand{\dif}{{\rm d}}

\newcommand{\A}{{\cal A}}
\newcommand{\C}{{\cal C}}

\newcommand{\bM}{{\bf M}}

\newcommand{\bq}{{\bf q}}

\newcommand{\br}{{\bf r}}

\draft
\begin{document}
\title{Integrability and Disorder in Mesoscopic Systems: \\
       Application to Orbital Magnetism}
\vspace{-1cm}

\author{ Klaus Richter$^{(1)}$, Denis Ullmo$^{(2)\ast}$,
and Rodolfo A.~Jalabert$^{(3)}$}
\address{$^{(1)}$Institut f\"{u}r Physik, Memminger Stra\ss e.
6, 86135 Augsburg, Germany and}
\address{ Max--Planck--Institut f\"ur Physik
komplexer Systeme, 01187 Dresden, Germany}
\address{$^{(2)}$Bell Laboratories, Lucent Technologies,
 1D-265, 600 Mountain Avenue, Murray Hill, New Jersey 07974-0636}
\address{$^{(3)}$Universit\'e Louis Pasteur, IPCMS-GEMME,
23 rue du Loess, 67037 Strasbourg Cedex, France}

\date{\today}

\maketitle
\vspace{-1cm}

\begin{abstract}
\vspace{-1cm}

We present a semiclassical theory of weak disorder effects in 
small structures and apply it to the magnetic response of 
non--interacting electrons confined in integrable geometries.
We discuss the various averaging procedures describing different
experimental situations in terms of one-- and two--particle
Green functions. We demonstrate that the anomalously large
zero-field susceptibility characteristic of clean integrable
structures is only weakly suppressed by disorder. This 
damping depends on the ratio of the typical size of the
structure with the two characteristic length scales describing
the disorder (elastic mean-free-path and correlation length of
the potential) in a power-law form for the experimentally
relevant parameter region. We establish the comparison with
the available experimental data and we extend the study of
the interplay between disorder and integrability to finite
magnetic fields.

\end{abstract}
\vspace{5mm}

\hspace{1cm} Invited paper for the Journal of Math.~Physics
\vspace{5mm}


\newpage


\section {Introduction}
\label{sec:Intro}

Electronic mesoscopic systems offer nowadays the 
possibility of being used as a laboratory for studying
quantum chaos. The main question of this novel
discipline -- the quantum signatures of the underlying
classical dynamics -- can be addressed in microstructures
defined on high mobility semiconductor heterojunctions. 
This connection presents a considerable challenge to
experimentalists since it implies complicated fabrication
processes and delicate measurements. The challenge for 
theoreticians is not lesser since semiconductor
microstructures are very rich condensed matter systems
(involving effects of temperature, confinement, disorder,
electron-electron and electron-phonon interactions, etc)
where the applicability and validity of simple models
has to be clearly established.

Within the simple model of a particle-in-a-billiard, 
important differences have been predicted \cite{Chaost},
and later measured \cite{Chaos,Albert}, in the transport
through chaotic and integrable geometries. In the former
nearby trajectories diverge exponentially and periodic
orbits are usually isolated; the latter  are
characterized by having as many constants of motion in 
involution as degrees of freedom, and periodic orbits
are organized in families on invariant tori \cite{gutz_book}. 
Chaotic cavities exhibit a universal behavior for the 
conductance fluctuations and weak-localization,
characterized by a single scale. On the contrary, 
integrable cavities do not show generic behavior 
presenting more fine-scale fluctuations and a non-lorentzian
line-shape of the low-field magneto resistance. In the
case of thermodynamical properties like the
magnetic susceptibility, the differences between
chaotic and integrable billiards are more spectacular
since they involve an order-of-magnitude enhancement
of the low-field susceptibility of integrable 
geometries compared to that of chaotic ones \cite{URJ95,RUJ95,vO94}. 
Unlike the transport
problem, the predicted different behavior according
to the integrability of the underlying classical
mechanics has not been experimentally confirmed yet.

The residual disorder present in actual microstructures
plays a special role in the quantum chaos studies.
Indeed, any perturbing potential, such as
the one provided by the disorder, immediately breaks the
integrable character of the classical dynamics.
Since small amounts of
disorder are unavoidable in actual microstructures, the
question whether or not integrable behavior should be
observed, naturally arises. It is then of foremost
importance to establish if the differences between 
chaotic and integrable geometries persist  when we
go beyond the particle-in-a-box model. This interplay
between integrability and disorder is the main subject
of this paper.

We start by characterizing the disorder. One limiting
case is the absence of it, where the dynamics is determined
by the non-random confinement potential (particle-in-a-box or
{\em clean} models). On the other extreme we have the
{\em diffusive} limit where the electron motion is a random
walk between the impurities and the confining effects are
not important. The strength of the disorder in the diffusive
case is characterized by the transport mean free path $\lt$:
the mean distance over which the electron momentum is
randomized. When $\lt$ becomes of the order of the typical
size $a$ of the microstructure, confinement {\em and} disorder
are relevant. 
For $\lt > a$ we arrive at the {\em ballistic}
regime where electrons can traverse the structure with a 
small drift in their momentum (going along almost straight 
lines), and their dynamics is mainly given by the bounces
off the walls of the confining potential. In the ballistic
regime the underlying classical mechanics still depends on 
the geometry and we would like to understand the different 
role of disorder in integrable and chaotic geometries. 

For short range impurity potentials (as typically found
in metallic samples) the scattering is
isotropic ($s$-type) and the momentum is randomized after
each collision with an impurity. There is therefore only
one length scale, namely $l_T$, characterizing the disorder.
For smooth impurity potentials (as typically realized in 
high-mobility microstructures) the scattering is forward
directed and $\lt$ may be significantly larger than the
elastic mean free path $l$ associated to the total amplitude
diffracted by the disorder \cite{DSS}. 
The regime $\lt > a > l$ is particularly
interesting because it is ballistic (since the classical
mechanics is hardly affected by disorder), but the single
particle eigenstates are short lived. In a more technical
language that we will precise on the sequel, we have $l$ as
given by a single-particle Green function and $\lt$ as by a
two-particle Green function \cite{AGD}. 
We will study the interplay 
between disorder and confinement for physical observables
that depend on one-- and two--particle Green functions,
concentrating on the magnetic susceptibility of
individual and ensembles of ballistic microstructures.

The natural tools to attack the interplay between disorder
and confinement are  semiclassical expansions since they
transparently convey at the quantum level the information
about the classical mechanics. Supersymmetry \cite{Efe83} 
and random matrix theories are quite powerful methods that 
have been widely used in recent studies of quantum chaos and 
disordered systems \cite{BarMel,JPB,Weid94,Efe95}, 
but are not applicable to our regime of interest
since they deal with the ergodic universal (long time)
properties of completely chaotic systems. Diagrammatic
perturbation theory for the disorder can 
describe the diffusive regime \cite{AltShk}, but calculations 
become exceedingly complicated when the confinement and the 
detailed nature of the impurity potential has to be 
considered. 

In our semiclassical approach we emphasize the dependence 
of disorder effects on the ratio between the finite
system size $a$ and the disorder correlation length $\xi$, showing that
confined systems exhibit strong deviations from the bulk--behavior.
In particular we demonstrate that for integrable geometries
the effect of smooth disorder results in a power--law 
damping of the two--particle Green function properties,
and we compare this behavior with that expected in chaotic systems.
For completeness of the presentation we first briefly review in
Sec.~\ref{sec:clean} our work on the magnetic response of
clean systems \cite{URJ95,RUJ95}.  We then develop in detail 
a treatment of disorder in ballistic microstructures extending some
preliminary work \cite{RapComm}. In Sec.~\ref{sec:disord}
we present the disorder model and some general implications at the level
of one-- and two--particle Green functions. In Sec.~\ref{sec:applic1} and 
\ref{sec:applic2} we specialize in the impurity averaged
magnetic susceptibility for individual and ensembles of 
microstructures.


\section {Orbital Magnetism in Clean Systems: A brief Review}
\label{sec:clean}

\subsection{Thermodynamic Formalism}
\label{subsec:thermo}

In this section we present the basic thermodynamical formalism for
obtaining the orbital magnetism within a semiclassical approach.
We indicate the main ideas for its application to clean microstructures
\cite{URJ95,RUJ95} which will be further developed in the Sections
\ref{sec:applic1},\ref{sec:applic2} in order to allow the treatment of
static disorder. The principle is to derive thermodynamical expressions for
the free energy and the grand potential using a semiclassical 
approximation for the density of states. This allows us to calculate
physical observables such as the magnetic susceptibility for the canonical
and grand canonical ensembles.

For a system of electrons confined to an 
area $A$ at temperature $T$ and subject to a perpendicular magnetic field $H$,  
the free energy $F(T,H,\bf N)$ for a fixed number $\bf N$ of
electrons and the grand potential $\Omega(T,H,\mu)$ (representing
the coupling to a particle reservoir with chemical potential $\mu$) 
are related by means of the Legendre transform
	\begin{equation} \label{eq:free}
	F(T,H,{\bf N}) = \mu {\bf N} + \Omega(T,H,\mu) \ .
	\end{equation}
The canonical $(\chi$) and grand canonical $(\cgc)$
susceptibilities of a confined electron gas are given by
	\begin{equation} \label{eq:susc_gc}
	\chi = - \frac{1}{A}
	\left(\frac{\partial^{2}F}{\partial H^{2}} \right)_{T,{\bf N}}
   \hspace{1cm} , \hspace{1cm} 
	\cgc = 	- \frac{1}{A}
   \left(\frac{\partial^{2}\Omega}{\partial H^{2}} \right)_{T,\mu} 
	\ .
        \end{equation}
The grand potential can be expressed in the form
\be
\label{eq:gcpot}
\Omega(T,H,\mu) = -\frac{1}{\beta} \int\,  {\rm d}E \, d(E) \ln[1 +
\exp(\beta(\mu-E))]
\ee
(with $\beta = 1/\kb T$) in terms of the single--particle density 
of states $d(E)$ which we decompose into a smooth mean and oscillating 
part according to
	\begin{equation} \label{eq:dosc1}
	d(E) = \bar d(E) + \dosc(E) \; .
	\end{equation}
As has been first noticed in the context of persistent currents
in disordered rings \cite{ensemble} a distinction between $\chi$
and $\cgc$ may be of crucial importance in mesoscopic thermodynamics: 
Although the number of electrons can be large for a mesoscopic system, 
the fact that $\bf N$ is fixed must be taken into account 
(by working in the canonical formalism) 
if a disorder or energy averaged magnetic response of an {\em ensemble}
of isolated micro-systems is examined. According to Imry\cite{Imrymeso} 
a convenient representation for the canonical free energy in terms of 
grand canonical quantities is obtained by 
expanding the relationship (\ref{eq:free}) to 
second order in $\mu-\bar{\mu}$ with a mean chemical potential
$\bar{\mu}$ being implicitly defined by accommodating $\bf N$ 
charge carriers to the mean number of states
	\be
{\bf N} = N(\mu) = \bar{N}(\bar{\mu})\ .
	\label{eq:mGCE}
	\ee
Here
\be
N(\mu) = \int_0^{\infty} \dif E \ d(E) \ f(E\!-\!\mu)  
\label{eq:NOS}
\ee
with the Fermi distribution function
	\begin{equation} \label{eq:fermi}
	f(E-\mu) = \frac{1}{1 + \exp [\beta (E-\mu)]} \, .
	\end{equation}
$\bar{N}$ is obtained in Eq.~(\ref{eq:mGCE}) by replacing 
$d(E)$ by $\bar{d}(E)$.
This finally allows  an expansion of the free energy as
\cite{ensemble}
	\begin{equation} \label{eq:fd}
	F({\bf N}) \simeq F^{0} + \Delta F^{(1)} +\Delta F^{(2)} \ ,
	\end{equation}
with
   \begin{mathletters}
   \label{allDF}
	\begin{eqnarray}
	     F^{0} & = & \bar \mu {\bf N} + \bar \Omega(\bar \mu) \; ,
	      \label{eq:DF0}  \\
	     \Delta F^{(1)} & = & \Oosc (\bar \mu)  \; ,
	        \label{eq:DF1} \\
	\displaystyle 
	     \Delta F^{(2)} & = & \frac{1}{2 \bar d (\bar \mu)} \
				\left( \Nosc (\bar \mu) \right)^{2} \ .
	     \label{eq:DF2}
	\end{eqnarray}
   \end{mathletters}
The functions $\Oosc(\bar \mu)$
and $\Nosc(\bar \mu)$ are expressed by means of Eqs.~(\ref{eq:gcpot}) and
(\ref{eq:NOS}), respectively, upon inserting the oscillating part $\dosc(E)$ 
of the density of states (\ref{eq:dosc1}).
The leading order contribution to $F$ is given by the first two
terms $F^{0} + \Delta F^{(1)}$ yielding the susceptibility 
calculated in the {\em grand} canonical case. $F^0$ gives rise to the
(two--dimensional) diamagnetic Landau--susceptibility which for
billiard like systems is expressed as for the bulk as
	\be
	\label{eq:chilandau}
	-\cl = -\frac{\gs e^2}{24 \pi m c^2} 
	\ee
with $\gs=2$ the spin degeneracy.

\subsection{Semiclassical treatment of susceptibilities}
\label{subsec:sus}

For a semiclassical computation of $\Delta F^{(1)}$ and $\Delta F^{(2)}$ 
and their derivatives with respect to $H$ we calculate $\dosc(E,H)$ from 
the trace 
	\be
	\label{eq:trace}
	d(E,H) = -\frac{\gs}{\pi} {\rm Im} \int d {\bf r} G_E({\bf r,r})
	\ee
of the semiclassical one--particle Green function. 
Its contribution to $\dosc(E)$ is given by \cite{gutz_book}
	\begin{equation}
	G_E({\bf r}',{\bf r}) = \sum_t D_t \exp{\left[i \left(
	\frac{S_t}{\hbar} - \eta_t\frac{\pi}{2}\right)\right]}
	 \ ,
	\label{eq:green}
	\end{equation}
as the sum over all classical paths $t$ (of non--zero length)
joining $\br$ to $\br'$ at energy $E$. 
	\be
	S_t = \int_{{\cal C}_t} {\bf p} \, d{\bf q}  
	\label{eq:cleanaction}
	\ee
is the classical action integral along the path ${\cal C}_t$. 
The amplitude $D_t$ takes care of the classical probability 
conservation, and $\eta_t$ is the Maslov index. 

The evaluation of the trace integral (\ref{eq:trace}) for chaotic
and integrable systems leads to the Gutzwiller-- \cite{gutz_book}
and Berry--Tabor-- \cite{ber76} periodic--orbit 
trace formulas, respectively. In order
to calculate the magnetic susceptibility at small fields one has to
carefully distinguish \cite{RUJ95} between the three possibilities 
of a chaotic billiard, the special case of an integrable billiard 
remaining integrable upon inclusion of the $H$--field, and the more
general case where the field acts as a perturbation breaking the 
integrability of a regular structure. 

Since our main interest in the Sections \ref{sec:disord},\ref{sec:applic1}
and \ref{sec:applic2}
will be devoted to disorder effects on the susceptibility of 
billiards being integrable at zero $H$--field we will focus here 
on the last case. There neither Gutzwiller-- nor 
Berry--Tabor--trace formulas are directly applicable and, following
Ozorio de Almeida \cite{ozo86},  a uniform treatment
of the perturbing $H$--field is necessary. In the integrable
zero--field limit each closed trajectory belongs to a torus
$I_\bM$ and we can replace $\br$ in the trace integral (\ref{eq:trace})
by angle coordinates $\Theta_1$ specifying the trajectory within
the (one--parameter) family and by the position $\Theta_2$ on the
trajectory. For small magnetic field the classical orbits can be 
treated as essentially unaffected while the field acts merely on the 
phases in the Green function in terms of the magnetic 
flux through the area ${\cal A}_\bM
(\Theta_1)$ enclosed by each orbit of family $\bM$. Evaluating the trace
integral (\ref{eq:trace}) along $\Theta_2$ for the semiclassical Green 
function of an integrable system leads in this approximation
to a factorization  of the density of states
	\be \label{eq:uniform_d}
        \dosc(E) = \sum_{\bsM \neq 0} \C_\bsM(H) \ d^0_\bsM(E)  
	\ee
into the contribution from the integrable zero--field limit
	\be
	\label{eq:d0}
	d^0_\bsM(E) = \tilde{D}_\bM \, \cos\left( \kf L_\bM - 
	  \eta_\bM \frac{\pi}{2} - \frac{\pi}{4} \right)
	\ee
($L_\bM$ is the length of the orbits of family $\bM$ and $\tilde{D}_\bM$
the semiclassical weight \cite{ber76}) and the function
	\be \label{eq:thetatrace} 
	\C_\bsM(H) = \frac{1}{2\pi} \int_0^{2\pi} \dif \Theta_1 \,
	  \cos \left[ 2 \pi \frac{H \A_\bsM(\Theta_1) }{ \Phi_0} \right] \;
	\ee
containing the $H$--field dependence ($\Phi_0 = hc/e$). Calculating 
$\Delta F^{(1)}$ from Eq.~(\ref{eq:DF1}) and taking the derivatives 
with respect to $H$ gives the grand canonical contribution to the 
susceptibility at small magnetic field
	\be \label{eq:chi1}
	\frac{\chi^{(1)}}{\cl} =- \frac{24\pi}{\gs}  m A 
	\left( \frac{\Phi_0}{2\pi A} \right)^2 \, 
	\sum_{\bM} \frac{R_T(\tau_\bsM)}{\tau_\bsM^2} \, 
        d^0_\bsM(\mu) \, 
        \frac{d^2 \C_\bsM}{d H^2} \, .
	\end{equation}
Here, $\tau_\bM$ is the period of a closed orbit of family $\bM$ and
\be
\label{eq:R_T}
R_T(\tau) = \frac{\tau/\tau_c}{\sinh(\tau/\tau_c)} 
\vspace{1cm} ; \hspace{1cm} \tau_c = \frac{\hbar \beta}{\pi}
\ee
is a temperature damping factor which arises from the convolution
integral in Eq.~(\ref{eq:gcpot})
and gives an exponential suppression of long orbits.
This is important from a physical as well as computational point
of view, as conceptual difficulties associated with the questions
of absolute convergence of semiclassical expansions at zero temperature 
do not arise.

Eq.~(\ref{eq:chi1}) is the basic equation for the susceptibility
of an individual microstructure.   When considering ensembles of 
ballistic microstructures  however, an average $(\overline{~\cdot~})$ 
over energy (i.e.~$\kf$) or over the system size $a$
 has  usually to be performed and leads to variations in the phases 
(actions $S/\hbar = \kf L_\bM$) of the density of states  (\ref{eq:d0})
which are much larger than $2\pi$. Therefore,
$\chi^{(1)}$ vanishes upon ensemble average. In order to
characterize the orbital magnetism of ensembles we introduce the
{\em typical} susceptibility $\chit = (\overline{\chi^2})^{1/2}$
(the width of the distribution) and the ensemble average 
$\overline{\chi}$ (its mean value, which is non--zero
because of the term $\Delta F^{(2)}$ in the expansion 
Eq.~(\ref{eq:fd})).   They are of theoretical
interest because of being based on two--particle Green functions
and they are relevant for the description of experiments on 
ensembles of mesoscopic systems.

If we assume that there are no degeneracies in the lengths of 
orbits from different families $\bM$ we obtain for $\chit$
	\be \label{eq:chit}
	\left( \frac{\chit}{\cl} \right)^2 = 
        \left(\frac{24 \pi}{\gs} m A\right)^2 
	\left( \frac{\Phi_0}{2\pi A} \right)^4 \, 
	\sum_{\bM} \frac{R_T^2(\tau_\bsM)}{\tau_\bsM^4}   \, 
	\overline{d^0_\bsM(\mu)^2}  \, 
        \left(\frac{d^2 \C_\bsM}{d H^2}\right)^2 \, .
	\end{equation}
In calculating $\overline{\chi}$, the grand canonical 
contribution $\chi^{(1)}$ from $\Delta F^{(1)}$ 
vanishes under energy average and the canonical correction 
$\Delta F^{(2)}$ in Eq.~(\ref{eq:fd}) gives in semiclassical 
approximation using Eq.~(\ref{eq:DF2}) 
	\be 
        \label{eq:chi2}
	\frac{\overline{ \chi}}{\cl}   \simeq
	\frac{\overline{ \chi^{(2)} }}{\cl}  = 
	- \frac{24 \pi^2}{\gs^2} \hbar^2
	 \left( \frac{\Phi_0}{2\pi A} \right)^2\, 
	\, \sum_{\bM} \frac{R^2_T(\tau_\bsM)}{\tau_\bsM^2} \, 
        \overline{ d^0_\bsM(\mu)^2} \, 
	\frac{d^2 \C^2_\bsM}{d H^2} \, .
        \ee
The Eqs.~(\ref{eq:chi1})--(\ref{eq:chi2}) provide the
general starting point for a computation of the susceptibility
of integrable billiards at small fields.

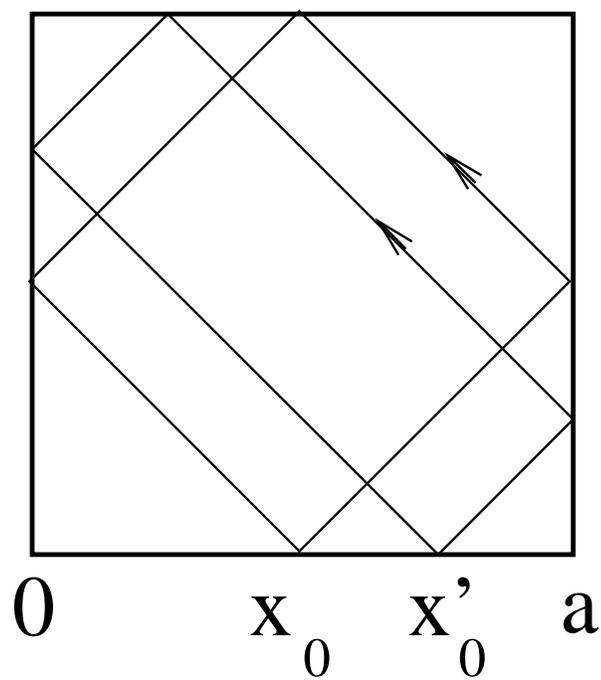
\begin{figure}
\caption{Two representative periodic orbits characterized by
         $x_0$ and $x_0'$ belonging to the family ${\bf M} =$ 
         (1,1) (denoting, respectively, one bounce with each
         wall) of a square billiard of length $a$. }
\label{fig:square}
\end{figure}

As an important example, which is also of experimental relevance
\cite{levy}, we will apply the results to square billiards. At
finite temperature $\chi$ is essentially given by the family
${\bf M} =$ (1,1) of the shortest, 
flux--enclosing periodic orbits depicted
in Fig.~\ref{fig:square}. A complete treatment
including families of longer orbits is given in Ref.~\cite{RUJ95}. 
Instead of $\Theta_1$ we use the lower reflection point 
$x_0$ as orbit parameterization within the family.
The orbits (1,1) have the unique length $L_{11} = 2 \sqrt{2} a$ and
enclose a normalized area $\A(x_0)=4 \pi x_0(a-x_0)/a^2$. 
Computation of $d_{11}^0 (\mu)$ for the square geometry gives
for $\chi^{(1)}$ (Eq.~(\ref{eq:chi1}))
\be \label{eq:chi1_sq}
        \frac{\chi^{(1)}}{\chi^0} =  
	\int_{0}^{a} \frac {{\rm d}x_0}{a} \A^2(x_0) 
	\cos (\varphi \A(x_0))
	\sin{\left(k_{\scriptscriptstyle F}
  L_{11}+\frac{\pi}{4}\right)} 
        \end{equation}
as a function of the  total flux $\varphi = H a^2/\Phi_0$ with
$\Phi_0 = h c/e$. The prefactor
	\be
	\label{eq:chi0}
	\chi^0 = \chi_{\scriptscriptstyle L}  
	\frac{3}{(\sqrt{2}\pi)^{5/2}} \, 
	(k_{\scriptscriptstyle F} a)^{3/2} R_T(L_{11}) \,  .
	\ee
shows the $(\kf a)^{3/2}$--dependence typical for (nearly--)integrable
systems.

For the square geometry the Equations (\ref{eq:chit}) and 
(\ref{eq:chi2}) for the susceptibilities $\chit$ and $\overline{\chi}$
(characterizing different ensemble averages) can be reduced to (including
only the dominant contributions from the family (1,1))
	\begin{equation}  \label{eq:chit_sq} 
	\frac{\chi^{(t)}}{\chi^0}   \simeq  
	\frac{\sqrt{\overline{{\chi^{(1)}}^2}}}{\chi^0} 
         = \frac{1}{\sqrt{2}} \  
        \int_{0}^{a} \frac{{\rm d}x_0}{a} 
	\A^2(x_0) 
	\cos \left( \varphi\A(x_0) \right)
	\end{equation}
and 
	\be 
        \label{eq:chi2_sq}
\frac{\overline{ \chi}}{{\overline{\chi}^0}} = \frac{1}{2}
        \int_{0}^{a} \frac{{\rm d}x_0}{a} \int_{0}^{a} 
	\frac{{\rm d}x_0'}{a}  \left[
	\A^{2}_{-} \cos(\varphi \A_{-}) +
	\A^{2}_{+} \cos(\varphi \A_{+})\right]
        \ee
with 
       \be
       \label{eq:chi0bar}
        \frac{\overline{\chi}^0}{\cl} = 
       \frac{3}{(\sqrt{2}\pi)^3} \ (\kf a) \  R^2_T(L_{11})
       \ee
and $\A_{\pm} = \A(x_0)\!\pm\!\A(x_0')$. Although the integrals   
(\ref{eq:chi1_sq}),(\ref{eq:chit_sq}) and (\ref{eq:chi2_sq}) can be
evaluated analytically in the clean case (leading to Fresnel 
functions of the magnetic flux\cite{URJ95}) the above expressions 
 serve as suitable starting points for the discussion 
of disorder effects on ensembles of microstructures
discussed in Sections \ref{sec:applic1} and \ref{sec:applic2}.


\section {Semiclassical approach to weak disorder}
\label{sec:disord}

Disorder is usually studied in terms of the ensemble average over
impurity realizations, since it is a perturbation of a electrostatic
potential whose detailed nature is unknown. Typically, quantum perturbation
theory is followed by the average over the strengths and positions of the
impurities. This approach is suited for macroscopic metallic samples
(which are self-averaging) or ensembles of mesoscopic samples (where
different samples present different impurity configurations). 
The possibility of measuring a single disordered mesoscopic sample 
posses a conceptual difficulty since there is not an average process 
involved. 
When discussing the effect of disorder on the orbital magnetism 
of microstructures, it is therefore necessary to distinguish 
between the behavior of an individual sample and an ensemble
\cite{AG95}.

Moreover, we have to consider the cases where the Fermi energy and
size of the microstructures are kept fixed under impurity average
and the cases where these parameters change with the different impurity
realizations. These various averages, that will be thoroughly discussed
in the remainder of the paper, can be expressed in terms of the
impurity average of one-- and two--particle Green functions. Therefore
we perform in this section  a general treatment of disorder effects
on the basis of semiclassical expansions of Green functions.  
The  Green function formalism, which is useful for
a wide range of physical problems, can 
be applied to thermodynamical quantities like the magnetic 
susceptibility (Sec.~\ref{sec:applic1} and \ref{sec:applic2}) 
as well as to quantum transport problems. 

\subsubsection{Disorder models}

Our basic assumptions for the treatment of disorder are the following: 
We study a spatially random potential $V(\br)$ characterized by a 
correlation function
	\be 
	\label{eq:corr}
	C(|\br-\br'|)  = \langle V(\br) V(\br') \rangle 
	\ee
 with a typical correlation length $\xi$ and a mean disorder
strength $C^0=C(0)$. We will make use of a Gaussian correlation 
	\be
	\label{eq:gauss_cor}
	C(|\br-\br'|) = C^0 \exp \left(-\frac{(\br-\br')^2}{4\xi^2}
                       \right)
	\ee
which allows us to derive analytical expressions for disorder averages 
considered below \cite{foot2}.
The disorder correlation function (\ref{eq:gauss_cor}) 
can be  viewed as being generated by means 
of a realization $i$ of a two-dimensional 
Gaussian disorder potential given by the sum

\be \label{eq:garapo}
V({\bf r}) = \sum_{j}^{N_i} \frac{u_j}
{2 \pi \xi^2} \ 
\exp{\left\{-\frac{({\bf r}\!-\!{\bf R}_j)^2}{2 \xi^2}
\right\}} 
\ee

\nin 
of the potentials of  $N_i$ {\em independent} 
impurities located at points ${\bf R}_j$ 
with uniform probability on an 
area ${\sf V}$. The strengths $u_j$ obey 
$\langle u_j u_{j \prime} \rangle = u^2 \delta_{jj'}$.
The disorder strength (as defined in Eq.~(\ref{eq:gauss_cor})) is
   \be
   \label{eq:C0}
C^0= \frac{u^2 n_i}{ 4\pi \xi^2}
   \ee
with $n_i = N_i/ {\sf V}$. For $\xi \rightarrow 0$ this model 
yields the white noise case of $\delta$-function scatterers
$V({\bf r})=\sum_{j}^{N_i} u_j \delta({\bf r}\!-\! {\bf R}_j)$.
We will use the model of Gaussian disorder for some analytical
calculations and for numerical quantum simulations. However, the
general results expressed in terms of the correlation function
$C(|{\bf r-r'}|)$ will be valid for any kind of disorder.

As we will show, disorder effects depend
on several length scales: the 
Fermi--wavelength $\lf$ of the electrons, the disorder correlation 
length $\xi$ and the size $a$ of the microstructure.
In the bulk case of an unconstrained two--dimensional electron 
gas (2DEG) we will distinguish between short range ($\xi\!<\!\lf$) 
and finite range ($\xi\!>\!\lf$) disorder potentials.
In the case of a microstructure a third, 
long range regime for $\xi\!>\!a\!>\!\lf$ has to be considered.
The cleanest samples used in
today experiments are in the finite range regime $a > \xi > \lf$ 
\cite{sweedish}.

\subsubsection{Single--particle Green function}

If we assume a microstructure with size $a\gg\lf$ (a condition 
which is always met in lithographically defined samples) 
and work in the finite
range or long range regime, where the disorder potential is 
smooth on the scale of $\lf$, a semiclassical treatment
is well justified. A natural starting point is the semiclassical
expression (\ref{eq:green}) for the single--particle Green 
function $G_E(\bf{ r',r})$ as a sum over the contributions from
classical paths. The classical mechanics of trajectories with
length $L_t \ll \lt$ (the transport mean free path) is essentially 
unaffected by disorder. Therefore the dominant effect on 
the Green function in Eq.~(\ref{eq:green}) results from
shifts in the semiclassical phases due to the modification 
of the actions while the amplitudes $D_t$ and topological 
indices $\eta_t$ are nearly unchanged. The first--order approximation
to the classical action (\ref{eq:cleanaction}) along a path 
${\cal C}_t$ in a system with weak disorder potential is
	\be 
	\label{eq:action_approx}
	S_t^d 
	\simeq S^c_t + \delta S_t \, ,
	\ee
where the clean action $S^c_t$ is obtained by integrating along
the {\em unperturbed} trajectory ${\cal C}_t^{\rm c}$ without disorder
(i.e.\ $S^c_t = \kf \, L_t$ in the case of billiards without magnetic field)
instead of the actual path ${\cal C}_t$. The correction term
$\delta S_t$ is obtained, after expanding ${\bf p} = \sqrt{2 m 
[E-V({\bf q})]}$ for small $V/E$, by the integral 
	\be 
	\label{eq:dis_action}
	\delta S_t \ = \ - \ \frac{1}{\vf} \ 
	\int_{{\cal C}_t^{\rm c}} V(\bq) \ {\rm d}q \ .
	\ee
In this approximation an impurity 
average $\langle \ldots \rangle$ acts only on $\delta S_t$
and the disorder averaged Green function reads
	\be \label{eq:avegre}
	\langle G_E({\bf r}',{\bf r})\rangle  =  \sum_t
	G_{E,t}^{c}({\bf r}',{\bf r}) \ 
	\langle \exp{\left[\frac{i}{\hbar} \delta S_t
	\right]} \rangle \ .
	\ee
Here $G_{E,t}^{c}$ is the contribution of the 
trajectory $t$ to the zero-disorder Green function $G_E^{c}$. 

For trajectories of length $L_t \gg \xi$ the 
contributions to $\delta S$ according to Eq.~(\ref{eq:dis_action})
from the disorder potential at trajectory
segments separated by a distance larger than $\xi$ are uncorrelated. 
The related stochastic accumulation of action along the path can be 
therefore interpreted as determined by a random-walk process, 
resulting in a Gaussian distribution of $\delta S_t(L_t)$. 
For larger $\xi$ or shorter trajectories ($L_t \not \gg \xi$), 
one can still think of a Gaussian distribution of the de-phasing $\delta S_t$
provided $V(\br)$ is generated by 
a sum of a large number of independent impurity potentials.
As a consequence of the Gaussian character of the distribution 
of $\delta S_t(L_t)$, the characteristic function involved in 
Eq.~(\ref{eq:avegre}) is given by
	\be
	\label{eq:gauss_av}
	\langle \exp{\left[\frac{i}{\hbar}\delta S_t\right]} \rangle =
	\exp{\left[-\frac{\langle \delta S^2_t \rangle}{2\hbar^2} 
	\right]}
	\ee
and therefore entirely specified by the variance 
	\be 
	\label{eq:dS2}
	\langle \delta S_t^2 \rangle = 
	\frac{1}{\vf^2} \int_{{\cal C}_t^c} {\rm d}q \ 
	  \int_{{\cal C}_t^c} {\rm d}q' \langle V(\bq) V(\bq') \rangle \ ,
	\ee
which is expressed as the mean of the disorder correlation function
$C(|\bq-\bq'|)$ when the unperturbed orbit is traversed. 

If we consider, to start with, an unconstrained 2DEG the sum in
Eq.~(\ref{eq:avegre}) is reduced to the direct trajectory
joining $\br$ and $\br'$.
If $L=|\br-\br'| \gg \xi$ the inner integral in Eq.~(\ref{eq:dS2}) 
can be extended to infinity and we obtain 
	\be
        \label{eq:dS2fr}
	\langle \delta S^2 \rangle = 
	   \frac{L}{\vf^2} \ \int {\rm d}q C(\bq) \; .
	\ee
The semiclassical average Green function for the bulk 
exhibits therefore an exponential behavior \cite{RapComm,Mirlin96}
(on a length scale $\lt > L \gg \xi$)
	\be \label{eq:avegrebulk}
	\langle G_{E}({\bf r}',{\bf r})\rangle \ = \
	G_{E}^{c}({\bf r}',{\bf r}) \ \exp{\left(-\frac{L}{2l}\right)} \, ,
	\ee 
 with the damping governed by an inverse {\em elastic} mean free path
	\be
	\label{eq:mfp}
	\frac{1}{l} = \frac{1}{\hbar^2\vf^2} \ \int {\rm d}q C(\bq) \, .
	\ee 
In the case of Gaussian correlation $C({\bf q})$ is given by 
Eq.~(\ref{eq:gauss_cor}) and we get 
        \be 
	\label{eq:mfp_C0}
       l= \frac{\hbar^2 \vf^2 }{ \xi \sqrt{\pi} C^0} \, .
        \ee   
Using the disorder strength (\ref{eq:C0}) we have 
        \be 
	\label{eq:mfp_gauss}
       l= \frac{4\sqrt{\pi}\hbar^2 \vf^2\xi }{u^2 n_i} \, .
        \ee   

In Appendix \ref{app1} we discuss the relation between the semiclassical
elastic MFP's (Eqs.~(\ref{eq:mfp})--(\ref{eq:mfp_gauss}))
and the MFP obtained from 
quantum diagrammatic perturbation theory for the bulk for the 
disorder model (\ref{eq:garapo}). The semiclassical
and the quantum result (Eq.~(\ref{eq:xi_expan})) 
agree asymptotically to leading order in $k_{\rm F}\xi$.
In the limit of small $\xi$, especially $\xi < \lambda_F$, 
our semiclassical approach is no longer applicable
\cite{schmid}. 
However, Eq.~(\ref{eq:avegrebulk}) still holds, but with $l$ replaced by 
$l_\delta$ given in Eq.~(\ref{eq:sigma_d}).

We now turn from the semiclassical treatment of the bulk to that of a
confined system. In the constrained case in the limit $\lt \ll a$ 
impurity scattering is the dominant process\cite{disor}. This gives 
rise to diffusive motion, and thus there is no essential difference 
to the bulk for the damping of the Green function. 
We will treat the ballistic regime $\lt\! >\! a$ where both,
the confinement {\em and} the impurities have to be considered.
A treatment of $\lt$ in the appendix shows that for finite 
$\xi$ the transport MFP $l$ is considerably larger
than the elastic one and a ballistic treatment is therefore well
justified, even if $l$ is of the order of the system size.

In contrast to the bulk case a disorder averaged confined system is no 
longer translationally invariant and one has to impose in quantum 
calculations the correct boundary conditions of the geometry.
Confinement implies semi-classically that $G^c_E(\br',\br)$ is
given as a sum over all direct and multiply reflected 
paths connecting $\br$ and $\br'$; disorder modifies the
corresponding actions according to Eq.~(\ref{eq:dis_action}). 

In the regimes of short-- and finite--range scatterers, the 
damping of each contribution $\langle G_{E,t}\rangle$ to 
$\langle G_E \rangle$ is given, analogous to
the bulk expression (\ref{eq:avegrebulk}), (using Eq.~(\ref{eq:dS2fr})) by 
	\be \label{eq:avegrefr}
	\langle G_E({\bf r}',{\bf r})\rangle  =  \sum_t
	G_{E,t}^{c}({\bf r}',{\bf r}) \ 
	\exp\left(-\frac{L_t}{2l} \right)   \ .
	\ee
Here, $L$ is now replaced by the trajectory length $L_t > a \gg \xi$. 
This gives an individual damping $\exp(-L_t/2l)$
for each geometry--affected path contributing to $\langle G_E\rangle$.

In the long range regime and for $\xi \sim a$ the correlation integral 
(\ref{eq:dS2}) can no longer be approximated (as for $\xi \ll L_t$)
by $L\int_{-\infty}^{+\infty} dq C({\bf q})$ due to correlations 
across different sectors of an orbit (with distance smaller $\xi$).
Therefore the orbit--geometry enters into the correlation integral.
For $\xi \gg a$ we can however expand $C(|\br-\br'|)$ and obtain 
in the case of Gaussian disorder (up to first order in $\xi^{-2}$)
$C(|\br-\br'|  \simeq C^0 [1- (\br-\br')^2/(4 \xi^2) ]$.
In this approximation the integral (\ref{eq:dS2}) gives for the Green 
function damping an exponent
\be \label{eq:dS2lr}
\frac{\langle \delta S_t^2 \rangle}{2\hbar^2} =  
\frac{1}{4\sqrt{\pi}}\, 
\frac{L^2_{t}}{l\, \xi}  \left(1-\frac{1}{2} 
\frac{I_t}{\xi^2} \right) \; .
\ee
$I_t = (1/L_t) \int_{{\cal C}_t} {\bf r}^2(q) {\rm d}q $ can be regarded as
the ``moment of inertia'' of the unperturbed trajectory ${\cal C}_t$ with
respect to its ``center of mass'' $(1/L_t) \ \int_{{\cal C}_t} 
\br(q) {\rm d}q$.  Eq.~(\ref{eq:dS2lr}) shows that the damping in the 
long range regime depends  quadratically on $L_t$ (in contrast to
linear behavior in the finite range case or bulk). The length scale of 
damping is now given by the geometrical mean of the bulk MFP 
$l$ {\em and} $\xi$. The leading damping term does not depend on
the specific orbit geometry since it essentially reflects the 
fluctuation in the mean of the (smooth) potentials of different 
impurity configurations. Inclusion of higher powers of $\xi^{-2}$
leads to additional contributions from higher moments $\int_{{\cal C}_t} 
r^n(q) {\rm d}q $ on the RHS of Eq.~(\ref{eq:dS2lr}).

\subsubsection{Two-particle Green function}

Density correlation functions in general or the 
typical susceptibility  (Eq.~(\ref{eq:chit})) and ensemble averaged
susceptibility (Eq.~(\ref{eq:chi2})), which will be treated in the
subsequent sections, involve squares of the density of states. 
Writing the latter, Eq.~(\ref{eq:trace}), in terms of the difference 
between advanced and retarded Green functions $(G^+-G^-)$
we are left with products of one--particle Green functions. The
terms of most interest are the cross products $G^+(r,r') \ G^-(r,r') =
G^+(r,r') {G^+}^*(r',r)$, because they survive the energy average and 
are sensitive to changes in the magnetic field. 

Since in the non--interacting approach
we are using, the two--particle Green function factorizes into a product
of one--particle Green function \cite{Doniach} we will 
use the former as a synonym for the latter.
The semiclassical average for products of single--particle Green 
functions will be quantitatively performed for the susceptibility of
confined integrable systems in Section \ref{sec:applic2}, and we 
discuss here the underlying ideas for the general case.

Considering for instance the product $G(r_1,r_2) G^*(r'_1,r'_2)$,
the effect of the disorder potential can be taken into account
perturbatively for each realization of the disorder in the same
way as before by
Eqs.~(\ref{eq:action_approx})--(\ref{eq:dis_action}).
Using the same kind of argument, one can therefore write
the disorder average as a double sum
over the averaged contributions from trajectories $t$ and $t'$
          \begin{eqnarray}
        \langle G_E \ G_E^\ast \rangle & = & \sum_t \sum_{t'} \
        \langle G_{E,t} \ G^\ast_{E,t'} \rangle
        = \sum_t \sum_{t'} \  G^c_{E,t} \ G^{c\ast}_{E,t'} \langle
        e^{(i/\hbar)(\delta S_t-\delta S_{t'})} \rangle
        \label{eq:Gproduct}\\
        & = &
         \sum_t \sum_{t'} \  G^c_{E,t} \ G^{c\ast}_{E,t'}
        \exp \left[-\frac{\langle (\delta S_t-\delta S_{t'})^2\rangle}{
        2\hbar^2} \right] \; . \nonumber
          \end{eqnarray}
It is necessary here however to take into account the correlation
of the disorder potential between points on trajectories $t$ and
$t'$.  One limiting case for instance would be that
$t$ and $t'$ are either the same trajectory or the time reversal
one of each other.  In these cases their contribution acquires
exactly the same phase shift and
 $\langle G_{E,t} \ G^\ast_{E,t} \rangle = |G^c_{E,t}|^2$.
Within our approximation the diagonal contributions $t=t'$,
which e.g.\ are responsible for the classical part of the
conductivity, remain thus disorder--unaffected, since we
assume the trajectories have a length much smaller
than $l_T$. (A semiclassical consideration of these effects
for trajectories of length of the order of $l_T$ or
larger was performed in
Ref.~\cite{Mirlin96} for the bulk, giving a damping of the
two--point Green function on the scale of $\lt$.)
At the opposite extreme,  if trajectories $t,t'$ are completely
uncorrelated, i.e.,
for long trajectories in classical chaotic systems or trajectories
in integrable systems with a spatial distance larger than
$\xi$, the average in Eq.~(\ref{eq:Gproduct}) factorizes:
$\langle G_{E,t} \ G^\ast_{E,t'} \rangle  =
\langle G_{E,t}  \rangle \cdot \langle  G^\ast_{E,t'} \rangle$
and lead to single--particle damping behavior.

The double sum Eq.~(\ref{eq:Gproduct}) may  however
involve pairs of trajectories
 which stay within a distance of the order of  $\xi$
(as for nearby paths on a torus of an integrable system).
In this case the behavior of
$\langle G_{E,t} \ G^\ast_{E,t'} \rangle$ is more complicated and
depends of the confinement geometry of the system under
consideration.  As a simple illustration of the interplay between
disorder correlation and families of orbits, let us consider
for the case of the bulk the product of $G(r_1,r_2)$ joining
$r_1 = (0,0)$ to $r_2 = (L,0)$ with $G^*(r'_1,r'_2)$ joining
$r_1 = (0,y)$ to $r_2 = (L,y)$, with $L\gg\xi$ but
$y$ possibly of the order of $\xi$.  Introducing the function
        \be \label{eq:K}
        K(y) = \int_{-\infty}^{+\infty} C(x,y) \ {\rm d} x
        \ee
(for Gaussian correlations Eq.~(\ref{eq:gauss_cor}),
$K(y)/K(0)= \exp(-y^2/(4\xi^2))$), 
the variance of the de-phasing is obtained as
        \be
        \langle  (\delta S_t-\delta S_{t'})^2\rangle =
        2 L \frac{(K(0) - K(y))}{\vf^2}
        \ee
and therefore $\langle G_E \ G_E^\ast \rangle =   G^c_E \ G^{c\ast}_E
\tilde{f}(y)$ with
        \be
        \label{eq:ftilde}
 \tilde{f}(y) = \exp\left[-\frac{L}{l_e}\left(1 - \frac{K(y)}{K(0)}\right)
         \right] \; .
        \ee
The function $\tilde{f}(y)$ expresses in a very simple way that as
$y \to 0$, the effect of disorder disappears ($\tilde{f}(0) = 1$)
while for $y\gg\xi$ the function $\tilde{f}(y)$ behaves as the square of 
single particle Green function damping.


\section{Fixed--size impurity average of the magnetic susceptibility}
\label{sec:applic1}

We consider here a disorder average (which will henceforth be called a
fixed--size impurity average)
of an ensemble of structures for which the parameters of the
corresponding clean system (geometry, size, chemical potential)
remain fixed under the change of impurity realizations.
In Section~\ref{sec:applic2}, we will then treat the more realistic case
of the orbital magnetic response of a {\em combined} energy
(or size) and disorder average. 

As  shown in the previous section, 
averages over weak disorder exponentially damp,
but do not completely suppress oscillatory contributions 
(with phase $\kf L_t$) to the single--particle Green function 
from geometrical paths in confined systems.  An 
observable quantity dependent on these contributions is the disorder 
averaged susceptibility of an
ensemble of billiards of the same size or same clean--system
Fermi energy, which will be studied first.

We will treat regular billiards at zero or small magnetic fields, where
the integrability is approximately maintained and the density of states
has the $H$--dependence of the formulae 
(\ref{eq:uniform_d})--(\ref{eq:thetatrace}).
The general result for $\chi^{(1)}$, Eq.~(\ref{eq:chi1}), formally
persists with the replacement of ${\cal C}_{\rm M}$ by
    \be \label{eq:thetatr_dis} 
  \langle \C_\bsM(H) \rangle  = \frac{1}{2\pi} \int_0^{2\pi} 
       \dif \Theta_1 \, \cos \left[ 2 \pi \frac{H \A_\bsM(\Theta_1) 
     }{ \Phi_0} \right] \ \exp\left[-\frac{\langle(\delta S_{\rm M}
        (\Theta_1))^2 \rangle}{2\hbar^2}\right] \; ,
	\ee
where $\langle\delta S_{\rm M}^2(\Theta_1)\rangle$ 
is given by Eq.~(\ref{eq:dS2})
with the integrals performed along the orbits of the family $\bf M$
parameterized by $\Theta_1$. In the finite range case (if all orbits
of a family $\bf M$ are of the same length as in billiards) each
family exhibits a unique disorder damping giving a contribution
	\be \label{eq:chiM_dis}
\langle	\chi^{(1)}_{\rm M}\rangle = \chi^{(1)}_{\rm M} \cdot 
                              \exp\left(-\frac{\langle\delta 
       S_{\rm M}^2\rangle}{2\hbar^2}\right)
        \ee
to the ballistic susceptibility. $\chi_{\bf M}^{(1)}$ is the 
contribution of family ${\bf M}$ to the clean susceptibility 
(Eq.~(\ref{eq:chi1})) and $\langle\delta 
S_{\rm M}^2\rangle/2\hbar^2 = L_{\rm M}/2l$.

In the case of square billiards, where the dominant 
contribution stems from the family (1,1), we obtain, in analogy with
Eq.~(\ref{eq:chi1_sq}),  
\be \label{eq:chi1_sq_dis}
        \frac{\langle \chi \rangle }{\chi^0} \simeq
        \frac{\langle \chi^{(1)} \rangle }{\chi^0} =
	\int_{0}^{a} \frac {{\rm d}x_0}{a} \A^2(x_0) 
	\cos (\varphi \A(x_0))  \left\langle
	\sin{\left(k_{\scriptscriptstyle F}
  L_{11}+\frac{\pi}{4} +\frac{\delta S(x_0)}{\hbar} \right)} \right\rangle
        \ee
with $\chi^0$ given by Eq.~(\ref{eq:chi0}). For a square billiard
$\delta S(x_0)$ is independent of $x_0$ for the finite-- as well
as for the long--range regime since $I_{11} = a^2/12$ (entering  into
Eq.~(\ref{eq:dS2lr})) is the same for all orbits (11).
Therefore Eq.~(\ref{eq:chiM_dis}) with ${\bf M} = (1,1)$ holds 
for both limiting cases. In the same way as for the damping of the
one--particle Green function (Eq.~(\ref{eq:avegrefr})) we obtain 
for square billiards at finite temperature in the finite range
regime
	\be \label{eq:chi1_disfr}
\langle	\chi \rangle  \simeq 
\langle	\chi^{(1)}\rangle  =  \chi^{(1)}_{\rm cl}  \cdot 
          \exp\left(-\frac{L_{11}}{2l}\right)  \, ,
        \ee
where $\chi^{(1)}_{\rm cl}$ denotes the susceptibility of the
system without disorder.

In order to control the validity of our analytical semiclassical
approximations we performed numerical quantum calculations by
diagonalizing the Hamiltonian for non--interacting particles
in a square billiard subject to a uniform perpendicular magnetic field 
and a random disorder potential of the form of Eq.~(\ref{eq:garapo}).
For a given selected correlation length $\xi$ and quantum 
mechanically calculated elastic MFP $l_{\rm 
qm}$ and fixed Fermi momentum $\kf$ the product of the mean number 
of impurities per area and squared mean impurity potential, 
$n_i u^2$, is determined by Eqs.~(\ref{eq:qmfp},\ref{eq:sigma_d}).
We found that our numerical results are essentially independent of the 
choice of $n_i$ (with $u^2$ adjusted accordingly) for $n_i \geq 200$
and used this value for the calculations presented here. The
positions ${\bf R}_j$ of the impurities were chosen as independently
distributed and for the $u_j$ we used a box distribution. 

Each impurity configuration $\alpha$ has a self--averaging effect for an 
{\em individual} square billiard (for $\xi < a$) due to the differences
of the impurity potential $V_\alpha({\bf r })$ across the structure. 
In an {\em average} over an ensemble of square billiards,
differences in the mean impurity potential $\overline{V_\alpha}
= (1/a^2) \int {\rm d} {\bf r} \ V_\alpha({\bf r})$ (the integral is 
taken over the area of the billiard) between different squares
lead to an additional damping.  It is characterized by the variance
\begin{eqnarray}
\label{eq:var_Vbar}
\langle \overline{V}^2 \rangle & = &  \frac{u^2 n_i}{a^2 \eta^2}
        \left[ \eta \ {\rm erf}(\eta) + \frac{1}{\sqrt{\pi}}
        \ \left(e^{-\eta^2} - 1 \right) \right]^2 \ \ ; \ \ 
        \eta = \frac{a}{2\xi} \\
        & \longrightarrow & \frac{u^2 n_i }{4 \pi \xi^2}  \qquad \qquad 
        {\rm for} \ \ \xi/a \longrightarrow \infty \label{eq:Vlr} \\
        & \longrightarrow & \frac{u^2 n_i }{a^2} \qquad \qquad  
        {\rm for} \ \ \xi \longrightarrow 0  
\end{eqnarray}
In the limit of  $\xi \gg a$ 
our numerical calculations showed that the self--averaging effect
is negligible (since the impurity potential is essentially flat
across the square) and the clean susceptibility of an {\em individual}
structure remains practically unaffected by disorder. In this limit
variations in the mean potential $\overline{V}$ of an ensemble
(Eq.~(\ref{eq:Vlr})) dominate the damping. 
In the limit of white noise disorder,
 fluctuations in the mean $\overline{V}$ of different
samples play a minor role and self--averaging is the 
predominant process for an integrable system: 
In semi-classical terms different trajectories 
of a family of closed orbits are perturbed by white noise disorder in an 
uncorrelated manner. Therefore we do not observe considerable
differences between the susceptibility of a single disordered billiard
of integrable geometry and the corresponding ensemble for $\xi \ll a$.
In a chaotic billiard this self--averaging effect does not 
exist (for not too small $\xi$, see end of Sec.~\ref{sec:applic2})), 
since orbits are isolated.
Therefore distinct differences between an individual disordered 
sample and an ensemble of disordered billiards are expected.

To improve the statistics of our numerical ensemble average for
square billiards we 
performed an average over disorder configurations with the same 
mean $\overline{V}$ and in addition averaged over $\overline{V}$ 
according to Eq.~(\ref{eq:var_Vbar}) \cite{foot4}.
Fig.~\ref{fig:chi1} shows results of the numerical quantum 
simulations for the average susceptibility $\langle \chi \rangle$ 
of an ensemble of squares
with fixed size but different disorder realizations at a temperature
$\kb T = 3 \gs \Delta$, where $\Delta$ is the mean level spacing. The
characteristic oscillations in $\kf a$ show an interchange between 
para-- and diamagnetic behavior on a scale $\kf L_{11}$. 
This indicates that they are dominated by contributions from the 
shortest flux--enclosing orbits of the family (1,1) (according to 
Eqs.~(\ref{eq:chi1},\ref{eq:chi1_sq_dis})), as has been already 
shown for the {\em clean} case in Refs.~\cite{URJ95,RUJ95}. 
Fig.~\ref{fig:chi1} demonstrates the damping of the clean susceptibility
(dotted line) with decreasing elastic MFP $l/a=4,2,1,0.5$ for fixed
$\xi/a=0.1$ (which represents a typical disorder correlation length in 
experimental realizations). Variations in the mean 
$\overline{V}$ lead to a de-phasing of the oscillations in the
finite range case on a scale $(\delta k) a \sim (4\pi)^{1/4} \sqrt{\xi/l_{
\rm qm}(\xi)}$ which is, as discussed above, 
small compared to the self--averaging effect in this regime. 

Fig.~\ref{fig:chi1_xidep} depicts the quantitative comparison 
between numerical and analytical results: It shows the logarithm of
$\langle \chi \rangle$ normalized to the corresponding zero--disorder
susceptibility as a function of the inverse MFP
for different correlation lengths $\xi$. The semi-classically
predicted exponential damping (Eq.~(\ref{eq:chiM_dis})) is shown
as straight lines for the short range ($\xi \ll a$, 
Eq.~(\ref{eq:avegrebulk}), full 
line for $\xi=0$) and long range ($\xi > a$, Eq.~(\ref{eq:dS2lr}), 
dotted lines for $\xi/a=$4,2,1 from the top). 
The semiclassical predictions accurately agree with the corresponding 
quantum results (symbols) for $\xi/a = 4, 2, 1, 0$ 
and fail for intermediate values $\xi/a = 0.5, 0.2$ 
(squares and diamonds) which are off the 
range of validity of the approximations. 
The transition from self--averaging dominated ($\xi \rightarrow 0$)
suppression to damping according to fluctuations in the floor $
\overline{V}$ (for $\xi/a \rightarrow \infty)$ turns out to be 
non--monotonic.


\section{Combined impurity-- and energy--average of the susceptibility}
\label{sec:applic2}

In currently experimentally realizable structures disorder averages 
cannot be performed independently from size--averages since the
detailed features of the confining potential do not remain unchanged for
different impurity configurations. From the basic expressions
(\ref{eq:chi1_sq}) and (\ref{eq:chi1_sq_dis})
for the susceptibility we see that changes 
in size $a$ give rise to rapid variations 
in the phase $\kf a$ (on a quantum scale) and a much slower
secular variation through the geometrical factors $\A$. Thus, the effect
of small size variations is equivalent to an energy ($\kf$) average.
As discussed in Sec.~\ref{sec:clean} for the clean case,
variations in $\kf$ lead to vanishing $\chi^{(1)}$. Therefore we have
to use the typical and energy averaged susceptibilities 
(see Eqs.~(\ref{eq:chit}) and (\ref{eq:chi2}) for their definition
in the clean case). When disorder is introduced we must consider
energy-- {\em and} disorder averages. The typical susceptibility is
now defined by $ \chi^{(t)} = \langle \overline{\chi^2} \rangle^{1/2}$.
It applies to the case of repeated measurements on a given microstructure
when different impurity realizations (and simultaneous changes in $\kf$)
are obtained by some kind of perturbation (e.g.\  cycling to room temperature).
From now on we will reserve the term $ \chi^{(t)}_{\rm cl}$ for the clean
typical susceptibility $ (\overline{\chi^2})^{1/2}$. The energy and impurity
averaged susceptibility $\langle \overline{\chi} \rangle$ describes the
magnetic response of an ensemble of a large number of microstructures with
different impurity realizations and variations in size. This is the 
situation of the experiment of Ref.~\cite{levy} that we discuss in the sequel.

\subsection{Integrable systems: The square billiard}

The semiclassical results for $\chi^{(t)}$ and $\langle \overline{\chi} 
\rangle$ for a system of  integrable geometry are obtained in an
analogous way as we proceeded for 
$\langle \chi \rangle$ in Section \ref{sec:applic1}, that is
by including in the integral (\ref{eq:thetatrace})
for ${\cal C}_{\bf M}$ a $\Theta_1$--dependent disorder--induced
phase $\exp(i\delta S(\Theta_1)/\hbar)$ 
(see Eq.~(\ref{eq:thetatr_dis})). 
However, now we have to take the square of ${\cal C_{\bf M}}$ 
(respectively $\partial^2{\cal C_{\bf M}}/\partial H^2$) before the 
impurity average and cross correlations between different paths
$\Theta$ and $\Theta'$ on a torus {\bf M} or between different
tori have to be considered. We discuss this effect,
typical of integrable systems, for the case of a square billiard.
For sake of clarity we moreover assume a temperature range such
that only the contribution of the shortest closed orbit has to be taken
into account.
Instead of Eq.~(\ref{eq:chit_sq}) and (\ref{eq:chi2_sq}) which
hold for the clean case, the contribution of orbits of 
topology ${\bf M} = (1,1)$ for the typical susceptibility now reads:
	\begin{equation}  \label{eq:chit_dis} 
	\left( \frac{\chi^{(t)}}{\chi^0} \right)^2  = \frac{1}{2} \  
        \int_{0}^{a} \frac{{\rm d}x_0}{a} 
	\int_{0}^{a} \frac{{\rm d}x_0'}{a}  \
	\A^2(x_0) \A^2(x_0')
	\cos \left( \varphi\A(x_0) \right)
	\cos \left( \varphi\A(x_0') \right)
	f(x_0,x_0')   \, ,
        \end{equation}
with $\chi^0$ defined as in Eq.~(\ref{eq:chi0}). The function
\begin{eqnarray}
 \label{eq:aveexpoad}
f(x_0,x_0') & = &
 \left\langle \exp{\left\{\frac{i}{\hbar}
\left(\delta S(x_0)-\delta S(x_0') \right) \right\}} \right\rangle \\
& = &  \exp{\left\{-\frac{1}{2\hbar^2} \left[\langle
\delta S^2(x_0)\rangle+\langle\delta S^2(x_0')\rangle - 2
\langle \delta S(x_0)\delta S(x_0')\rangle \right]
 \right\}}
\end{eqnarray}
accounts for the effect of disorder on pairs of orbits
$x_0$ and $x_0'$. See Eq.~(\ref{eq:ftilde}) for the treatment in the
general case. For the magnetic response of an 
energy-- and disorder--averaged ensemble we find correspondingly:
        \begin{equation}
        \label{eq:chia_dis}
\frac{\langle\overline{ \chi} \rangle}{{\overline{\chi}^0}} =
        \frac{1}{2} \ \int_{0}^{a} \frac{{\rm d}x_0}{a} \int_{0}^{a} 
	\frac{{\rm d}x_0'}{a}  \left[
	\A^{2}_{-} \cos(\varphi \A_{-}) +
	\A^{2}_{+} \cos(\varphi \A_{+})\right]
	f(x_0,x_0') 
	\end{equation}
with $\overline{\chi}^0$ defined in Eq.~(\ref{eq:chi0bar}) 
and $\A_{\pm}$ as in Eq.~(\ref{eq:chi2_sq}).

\subsubsection*{1. Short range case}

We begin with the discussion of the short range case:
Although we reach the border of applicability of  our semiclassical
approximation for $\xi \longrightarrow 0$, it shows us that in this
limit orbits with $x_0 \neq x_0'$ are disorder--uncorrelated 
and all such pair contributions are exponentially damped.
Using exclusively the family (1,1), one obtains
 an overall suppression of the typical and
average susceptibility at finite temperature according to
	\begin{eqnarray}
	\label{eq:xi0t_dis}
	\lim_{\xi \rightarrow 0} \chi^{(t)} 
	         & = & \chi^{(t)}_{\rm cl} \ e^{-L_{11}/2l_\delta}  \\
	\label{eq:xi0a_dis}
	\lim_{\xi \rightarrow 0} \langle\overline{\chi} \rangle
	         & = & \overline{\chi} \ e^{-L_{11}/l_\delta}   \, .
	\end{eqnarray}
Note that the exponent for $  \langle\overline{\chi} \rangle $
differs by a factor $1/2$ from that for
$\langle \chi \rangle$ (see Eq.~(\ref{eq:chiM_dis}) and subsequent text). 

Fig.~\ref{fig:chia_xi0} depicts the $\kf a$ dependence of the
ensemble averaged susceptibility $\langle \overline{\chi}
\rangle$ in the short range case $\xi=0$. The dotted curves
showing the semiclassical analytical formula (\ref{eq:xi0a_dis})
are compared with a direct quantum mechanical calculation of
$\langle \chi^{(2)} \rangle$ (using the numerically obtained 
$\Nosc (\bar \mu)$ in Eq.~(\ref{eq:DF2}))
for disorder ensembles of different impurity strength
equivalent to an elastic MFP $l_\delta/a=\infty, 8, 4, 2$ and 1 
at $\kf a \sim 65$ (from the top). Note, that the effective MFP decreases 
along the curves with decreasing $\kf$ (see Eq.~(\ref{eq:sigma_d})) 
and the localized regime may eventually be reached for small $\kf a$.
At the limit of the ballistic regime at small $l \sim a$ the 
semiclassical result begins to differ from the quantum one
although the functional behavior remains the same. This arising
difference may be related to non--ballistic scattering from impurities
which is not included here.

\subsubsection*{2. Finite range case}

In the finite range $\lf < \xi \ll a$, the phase shifts $\delta S(x_0)$ and
$\delta S(x_0')$ in $f(x_0,x_0')$ are accumulated in a correlated way,
if the spatial distance of two orbits $x_0$ and $x_0'$ is smaller than 
$\xi$. To evaluate the product term 
$2 \langle \delta S(x_0)\delta S(x_0')\rangle $ in the exponent
of $f(x_0,x_0')$ in this regime the
integrations are performed as in Eq.~(\ref{eq:dS2}) but with
$\bf q$ and $\bf q'$ running along paths starting at $x_0$,
respectively $x_0'$. 
Ignoring the additional correlations
occuring near the bounces off the boundaries of the billiard, the 
trajectories $x_0$ and $x'_0$ (see Fig.~\ref{fig:square})
can be regarded as straight lines
remaining at a constant distance $y=|x_0 - x'_0|/\sqrt{2}$
from another. We can therefore approximate $f(x_0,x'_0)$ by
$\tilde f(|x_0 - x'_0|/\sqrt{2})$ with the function $\tilde f$
given by Eq.~(\ref{eq:ftilde}).  For Gaussian correlation we thus have
	\begin{equation} \label{eq:f_gauss}
	f(x_0,x_0')  =   \exp\left\{-\frac{L_{11}}{l}
	\left[ 1 - \exp\left(-\frac{(x_0-x_0')^2}{8\xi^2}\right)\right]
	\right\}  \, .
	\end{equation}
Orbits separated by $|x_0-x_0'| \gg \xi$ are disorder--uncorrelated
and exponentially suppressed: 
$ f(x_0,x_0')  \simeq \exp(-L_{11}/l)$. For those
orbits the individual random disorder leads to an uncorrelated
detuning of the phases. In contrast to that, disorder 
only weakly affects trajectories separated by $|x_0-x_0'| < \xi$.


The disorder averages in the finite range regime lead,
 by means of the function $f$, to a non--exponential damping of
the susceptibilities for systems with families of periodic orbits. 
This behavior becomes obvious for the case of square billiards
where at $H=0$ the integrals (\ref{eq:chit_dis}) and (\ref{eq:chia_dis}) 
can be evaluated analytically in the limits of $L_{11} \ll l$ 
(extreme ballistic) and $L_{11} \gg l$ (deep ballistic).
We find for the typical and average susceptibility at $H=0$ in 
the finite range case for $L_{11} \ll l$ 
	\begin{mathletters} 
        \label{eq:eb}
        \be \label{ebt}
        \left(\frac{ \chi^{(t)} }{
        \chi^{(t)}_{\rm cl}} \right)^2 \simeq
	1-\frac{L_{11}}{l} \left(1-c_t\ \frac{\xi}{a} \right)  \ ,
	\ee
        \be \label{eba}
        \frac{\langle \overline{\chi}\rangle}{
         \overline{\chi}} \simeq
	1-\frac{L_{11}}{l} \left(1-c_a\ \frac{\xi}{a} \right)  \ ,
	\ee
	\end{mathletters}
and for $L_{11} \gg l$ (by steepest descent):
	\begin{mathletters} 
        \label{eq:db}
        \be \label{dbt}
        \left(\frac{ \chi^{(t)} }{
        \chi^{(t)}_{\rm cl}} \right)^2  \simeq 
	c_t \ \left(\frac{\xi}{a}\right) \ 
	\left(\frac{l}{L_{11}}\right)^{1/2}   \ ,
	\ee
	\be \label{dba}
        \frac{\langle \overline{\chi}\rangle}{
         \overline{\chi}}  \simeq 
	c_a \ \left(\frac{\xi}{a}\right) \ 
	\left(\frac{l}{L_{11}}\right)^{1/2} \ .
	\ee
	\end{mathletters}
The constants in the above equations are 
$ c_t = (20/7) \sqrt{2 \pi} $ and $c_a = 2 
\sqrt{2\pi}$.   Eqs.~(\ref{eq:eb}) express the limit of very
weak disorder, showing that the small disorder effect
is further reduced due to the correlation of
the disorder potential.
The other limit, Eqs.~(\ref{eq:db}), is noticeably more
interesting since it shows that disorder correlation effects
lead to a replacement of the exponential
disorder damping by a power law.

Fig.~\ref{fig:chiav} depicts in logarithmic representation
our collected results for the disorder averaged typical (a) 
and averaged (b) susceptibility for square
billiards (at $H=0$ and $\kb T = 2\gs\Delta$) as a function of
the inverse elastic MFP for different disorder correlation
lengths. The symbols denote results from numerical quantum 
simulations described in the previous section and the 
full curves semiclassical results from numerical integration
of the Eqs.~(\ref{eq:chit_dis}) and (\ref{eq:chia_dis}). For the
short range case $\xi = 0$ they reduce to Eq.~(\ref{eq:xi0a_dis})
predicting an exponential decrease with exponent $L_{11}/l$
which is in line with the quantum calculations (circles). The 
semiclassical results for the finite range are on the whole
in agreement with the numerical results for $\xi/a = 0.1$ (diamonds),
$\xi/a = 0.2$ (triangles) and $\xi/a=0.5$ (squares). The
semiclassical curves seem to overestimate the 
damping of the typical susceptibility. The dotted curves (shown for
$a/l \geq 1$) depict the analytical expressions (\ref{eq:db}) 
in the regime $L_{11} > l$. Since for finite 
$\xi$ the transport MFP $l_T > l$ (see Eq.~(\ref{eq:qtmfp})), 
this regime can still be considered as (deep) ballistic and 
our semiclassical assumptions being based on straight--line 
trajectories remain valid. 

As the semiclassical formulae already
indicate, the overall disorder behavior of $\langle \overline{\chi} 
\rangle$ and $ \chi^{(t)}$ is quite similar.

\subsubsection*{3. Long range case}

For completeness, we will consider the effect of the disorder
for the long range regime: We can use the Eqs.~(\ref{eq:chit_dis}) 
and (\ref{eq:chia_dis}) but cannot calculate
the disorder function $f(x_0,x'_0)$ in the same way as for the
finite range. We can however, similar as for
$\langle \chi \rangle$ in Section \ref{sec:applic1}, expand the
exponent $-\langle (\delta S(x_0)-\delta S(x_0'))^2 \rangle$
of $f(x_0,x_0')$ in Eq.~(\ref{eq:aveexpoad}) for small $a/\xi$.
In the case of the square all orders up to $(a/\xi)^8$ vanish
and we find a {\em very} small overall reduction of the clean averaged 
susceptibilities (from family (11)) given by 
        \be \label{eq:chitlr}
        \left(\frac{ \chi^{(t)} }{
        \chi^{(t)}_{\rm cl}} \right)^2 \simeq
   1 - 6.5 \cdot 10^{-5} \ \frac{a}{l}\ \left(\frac{a}{\xi}\right)^9 \, .
	\ee
For square billiards this leading order contribution does not depend 
any longer on $x_0$. 
The energy-- and disorder--average $\langle \overline{\chi} 
\rangle$ exhibits the same damping as $(\chi^{(t)})^2$. 
Note that besides the high order
in $(a/\xi)$ the prefactor is rather small. This weak suppression of the
averaged susceptibilities can be related to the fact that in the long
range case, different sectors of the contributing periodic orbits are
highly correlated. As visible in Fig.~\ref{fig:chiav}(a), the quantum
mechanical results (squares) for $\chi^{(t)}$ at $\xi/a = 0.5$, 
which are closest to the long range case, exhibit already a very weak damping.

\subsection{Disorder effects at finite $H$--field: from integrable to chaotic
behavior}

In Fig.~\ref{fig:Hdepend} we compare the ratio $(\chi^{(t)}/\chi^0)^2$ 
(obtained from calculating the integral in Eq.~(\ref{eq:chit_dis}))
as a function of the dimensionless 
flux $\varphi=Ha^2/\Phi_0$ for the clean case and
for disorder characterized by $l=a$ and $\xi = 0.1$. 
This figure shows that the damping due to disorder
is maximal at zero field, but that already for $\varphi=5$
the disorder seems not to affect the magnetic response any further.

The origin of this behavior can be understood readily by
observing that as soon as $\varphi$ is larger than one, the
integral Eq.~(\ref{eq:chit_dis}) is correctly approximated
by  a stationary phase approximation\cite{URJ95}.  The stationary point
$x_0^s = a/2$ corresponds to the two periodic orbits of the
{\em perturbed} system, and only the trajectories such that
	\be \label{eq:criter}
	 (x_0 - x_0^s)^2 \varphi < 1
	\ee
 actually contribute to the integral.
The magnetic field causes a de-phasing of the 
contributions of the various trajectories of the family,
thus breaking the integrability of the system.
This effect is responsible for
the overall decrease of the typical susceptibility as the field increases.
In this respect clean and disordered square billiards are not
equivalent. In the disordered case, trajectories
separated by a distance larger than $\xi$ are already  not
contributing in phase.  Therefore the additional
magnetic field affects the magnitude of the susceptibility much less.
This remains true up to the
point where the condition (\ref{eq:criter}) implies
$|x_0 - x_0^s| < \xi$ in which case the disorder is not
effective anymore, and the two curves coincide.

Therefore the behavior of the disorder damping we discussed 
in the previous subsection is characteristic for integrable geometries.  
For chaotic systems diagonal contributions (pair products of the
same periodic orbit) are barely affected
by disorder.  This behavior is similar to that of an integrable
systems at finite field. When evaluating the contribution to
the trace of the Green function in the neighborhood of
a periodic orbit by stationary phase approximation,
(as for the derivation of the Gutzwiller trace formula)
only  orbits extremely close to the periodic orbit under consideration
actually contribute.  Unless $\xi$ is taken exceedingly small,
all these trajectories will see the same disorder potential.

As  a final remark, note that {\em non--diagonal} contributions 
(pairs of different paths) are fully damped upon impurity average 
for chaotic as well as integrable systems, since the disorder potential
along two different trajectories is usually not correlated (see also the
discussion of the averaged Green function product after 
Eq.~(\ref{eq:Gproduct})).
Therefore non--diagonal contributions, which might be necessary
to be considered in the clean case\cite{foot6},
are exponentially suppressed in the presence of disorder. On the contrary,
{\em diagonal} terms which contain orbit correlations on 
distances $\xi$, exhibit non--exponential behavior
(Eq.~(\ref{eq:f_gauss})) as a function of the inverse MFP $1/l$
for integrable geometries and are not affected (within our 
approximations) by disorder in the chaotic case.


\subsection{Relation to experiment and other theories}
\label{subsec:exp}

Measurements of the orbital magnetism of small
microstructures are still rare today. The only experiment on
ensembles of ballistic billiards that we are aware of, was performed
by L\'evy et al. \cite{levy} and investigated the magnetic
susceptibility of an array of about $10^5$ ballistic square-like
cavities. The size of the squares is on average $a=4.5 \mu m$, with 
a large dispersion (estimated between 10 and 30\%) along the array.
Each individual square is a mesoscopic ballistic system since 
the phase-coherence length is estimated to be 
$L_{\Phi} = 15$--$40$ $\mu m$ and the elastic mean-free-path 
$l = 4.5$--$10$ $\mu m$. 
The potential correlation length can be estimated \cite{sweedish}
to be of the order of $\xi/a \simeq 0.1$. Taking the most
unfavourable case of $l \simeq a \simeq 4.5 \mu m$ we obtain,
with respect to the clean case, a disorder reduction 
for the averaged susceptibility of
$\langle \overline{\chi} \rangle /\overline{\chi} \simeq 0.37$,
showing that the features of the clean integrable systems (strong
paramagnetic susceptibility at $H=0$) persist upon inclusion
of disorder. Since $\overline{\chi} \simeq 100 \ \cl$ 
\cite{URJ95,RUJ95,vO94}, our calculations
for the paramagnetic response of the 
ballistic squares agree quantitatively with the experimental 
findings (given the experimental uncertainties). 

Persistent currents in individual quasi-ballistic rings
have recently been measured \cite{BenMailly}. 
A similar setup would be needed for measuring the magnetic
response of singly connected geometries, where our 
typical susceptibility (\ref{eq:chit_dis}) should be measured
for the integrable case. Since modern lithographic techniques
allow to design chaotic as well as integrable cavities 
\cite{Chaos,Albert} and since we have demonstrated that disorder
does not mask this difference, an order-of-magnitude effect is
expected in the susceptibility according to the
shape (chaotic vs.\ integrable) of the cavity.

In a related theoretical work
Gefen et al.\ \cite{braun} followed a complementary approach
to ours and calculated the disorder--averaged susceptibility 
for an ensemble of ballistic squares based on long trajectories
[strongly] affected by scattering from $\delta$--like impurities.
They found that the average susceptibility does not depend
on the elastic MFP. These results are not borne out by
either our analytical or our semiclassical calculations at
temperatures relevant for the experiment, where the exponential
damping from Eq.~(\ref{eq:R_T}) makes very long trajectories 
irrelevant.


\section {Summary}
\label{sec:summary}
In this work we have studied the interplay between integrability
and disorder in the ballistic regime. The integrable property 
of the confining potential of a microstructure implies a peculiar
behavior of its thermodynamical response functions, like the
magnetic susceptibility. The disorder effects provided by remote
impurity scattering tend to weaken the importance of the boundary
scattering (and therefore the relevance of the underlying classical
mechanics). Using a semiclassical approach we quantify this damping
and show it to be much weaker than previously estimated (power-law
suppression instead of exponential damping for the typical and 
average susceptibility). The disorder damping is decisively affected
by finite--size effects since it depends not only on bulk-like 
characteristics of the disorder (like the elastic mean-free-path),
but also on the ratio between the size of the structure and the
correlation length of the potential. 

Our finding for the weak disorder damping is particularly important
due to the large phase coherence effects found for clean 
integrable structures and to
the fact that the difference in the magnetic response 
between integrable and chaotic geometries
has not been experimentally demonstrated yet.

Our calculational tools have been semiclassical expansions, which
naturally convey at the quantum level the information about the
underlying classical mechanics and its sensitivity with respect to
disorder. For the weak disorder that we have considered in this work,
the lowest order approximation consists of the perturbative modification
of the classical actions by the impurity potential. Averages over
impurity configurations following our semiclassical calculations, 
allow us to obtain various ensemble susceptibilities. Our analytical
calculations have been checked against numerical quantum 
simulations performing exact diagonalizations 
of the corresponding Hamiltonian. 

The need to consider different averages is inherent to ballistic
nanostructures, which are sufficiently small to be non-self-averaging.
These various types of impurity-averaged susceptibilities for 
integrable systems are summarized in Table I for the three regimes 
defined by the correlation length of the impurity potential. We have
first studied the fixed--size averaged susceptibility, directly
obtainable from the disorder average of one-particle Green functions.
It corresponds to the case where different impurity configurations
of a given sample with a fixed Fermi energy are considered. For the short
range regime, where the disorder correlation length $\xi < \lf$,
we have an exponential suppression of the clean results governed
by the short-range elastic mean-free-path $l_{\delta}$ and the length
of the most relevant trajectories. This result also holds in the 
finite-range ($\lf < \xi \ll a$), 
but with an elastic mean-free-path that we have evaluated semi-classically. 
In the long-range regime ($\xi > a$) the fixed--size  averaged
susceptibility depends exponentially on the product $(L/l)\cdot(L/\xi)$
(where $L$ denotes the typical orbit length)
and a correction taking into account the geometry of the
periodic trajectories. 

For comparison with actual experiments we have to take into account
that different impurity realizations are obtained together with a change
in the Fermi energy and the size of the structures. We are then lead to 
consider impurity and size averaged susceptibilities, which are 
expressed in terms of two-particle Green functions. The typical
susceptibility is appropriate when considering the magnetic response of
an individual sample which is thermally cycled in order to obtain
different realizations of the potential. The average susceptibility is
obtained from the measurement of an array of microscopically different 
samples. For the short-range case the only difference between one--
and two-point Green function quantities is the factor $1/2$ of the
exponential damping of the former. In the finite-range regime there
appear important differences when considering two-point Green function
quantities with respect to the one-particle case. Closed trajectories
that remain closer than the correlation length of the potential result 
in a weak damping with a power-law dependence on $l/L$ and $\xi/a$.
This is the experimentally relevant situation, and the use of 
standard parameters lead us to conclude that disorder damping in
currently realizable microstructures is sufficiently weak in 
order not to mask the large effects due to integrability. 
In the long-range case the damping due to disorder is extremely small.

We have further considered the interplay between disorder and magnetic
field in integrable geometries. It is interesting to notice that both
have a similar effect since they produce de-phasing between nearby
trajectories. Since the two sources of de-phasing do not superpose,
we find that disorder is less effective at finite fields, and 
reciprocally, disordered samples are less sensitive to magnetic field. 

In chaotic geometries periodic trajectories are usually isolated,
resulting in smaller oscillations of the density of states and a
much smaller magnetic response than integrable systems. Introduction of
disorder in chaotic geometries is therefore less dramatic than in
integrable systems, since it merely changes the action of the relevant
periodic trajectories instead of producing de-phasing within a family.
The transition from the ballistic
regime (where classical trajectories are essentially unaffected by
disorder) to the diffusive regime will be considered in a subsequent
publication.

In this work we have started from a system that is physically
realizable using modern technology and we have developed a
theoretical model with some key ingredients involving integrability
and disorder. These are deep theoretical issues that need to be
complemented by the consideration of other effects, like interactions,
in order to obtain a complete description of the thermodynamics of
mesoscopic systems.  

\section*{Acknowledgements}

We would like to thank Harold Baranger for numerous discussions
and Paul Walker for a careful reading of the 
manuscript. KR and RAJ acknowledge support from the 
``Coop\'eration CNRS/DFG" (EB/EUR-94/41) and the French--German
program PROCOPE.


\appendix

\section {Relation between semiclassical and quantum mechanical results 
   for bulk mean free paths}
\label{app1}

It is instructive to compare the semiclassical results of 
Eqs.~(\ref{eq:mfp})--(\ref{eq:mfp_gauss}) for the ballistic regime
with their counterparts obtained from quantum mechanical scattering 
theory. 

In a perturbative diagrammatic approach (treating the related
Dyson--equation for scattering within 
a self--consistent Born approximation) 
the damping of the disorder--averaged one--particle Green 
function in a random potential is of the same 
exponential form as in Eq.~(\ref{eq:avegrebulk}) \cite{AGD}.
This is usually obtained by replacing the imaginary part of
the self-energy in the Green function after impurity average
by the density of states of the unperturbed system.
The resulting quantum mechanical inverse elastic MFP $l_{\rm qm}$,
which appears in Eq.~(\ref{eq:avegrebulk}), is related
to the total cross section $\sigma$ by means of
  \be
	\label{eq:mfp_cross}
	\frac{1}{l_{\rm qm}} = n_i  \sigma   \, ,
  \ee
where $n_i$ is the impurity density and 
  \be 
	\label{eq:cpart}
        \sigma = \int \ {\rm d} \Theta \ \sigma(\Theta)
  \ee
with $\sigma(\Theta)$ being the partial cross section for
scattering with an angle $\Theta$.
   
For a Gaussian disorder potential of the form of 
Eq.~(\ref{eq:garapo}) a calculation of the cross section
can be performed analytically and the corresponding inverse MFP  
gives
	\be
	\label{eq:qmfp}
	\frac{1}{l_{\rm qm}} = 
	\frac{1}{l_\delta} \, I_0[2(k\xi)^2] \, e^{-2 (k\xi)^2} \, .
	\ee
Here, $I_0$ is a modified Bessel function and 
	\be
	\label{eq:sigma_d}
	\frac{1}{l_\delta} 
              = \frac{2\pi}{\hbar} \frac{n_i u^2}{v_{\rm F}} d(\mu)
              =  \frac{n_i u^2}{v_{\rm F}} \frac{m}{\hbar^3} 
	\ee
is the inverse MFP for the white noise case of $\delta$--like 
scatterers of mean strength $u$.
The $v_{\rm F}$ is the Fermi velocity  and 
$d(\mu)=m/(2\pi\hbar^2)$ the density of states at the 
Fermi energy of a 2DEG \cite{AGD}.

In order to compare $l_{\rm qm}$ with our 
semiclassical result we expand
$l_{\rm qm}(k\xi)$ for large $k\xi$ which gives
	\be
	\label{eq:xi_expan}
l_{\rm qm}(k\xi)  \simeq
 \sqrt{4\pi} \ (k\xi) \ l_\delta \left[1-\frac{1}{16(k\xi)^2} \right]
\hspace{1cm} \ {\rm for} \qquad  k\xi \ \longrightarrow \infty \, .
\ee
The leading order term is exactly the semiclassical MFP 
Eq.~(\ref{eq:mfp}) for the Gaussian disorder model (\ref{eq:garapo}).
The agreement between the semiclassical and diagrammatic approaches for
the bulk can be related to the fact that our semiclassical
treatment of disorder corresponds to the use of the
Eikonal approximation (for each single scattering event) which 
is known to give the same results as Born approximation for large 
$k\xi$.

In the limit of $\xi < \lambda_F$ where our semiclassical description
is no longer applicable, the mean free path $l_{\rm qm}$ approaches 
$l_\delta$, which means that Eq.~(\ref{eq:avegrebulk}) can further be 
used, but with the semiclassical $l$ replaced by $l_\delta$.

The quantum mechanical transport mean free path $l_{\rm T}$ 
is calculated by including a factor $(1-\cos \Theta)$ 
in the integral (\ref{eq:cpart})
for the scattering amplitude. It reads for Gaussian disorder
\begin{eqnarray}
\label{eq:qtmfp}
\frac{1}{l_{\rm T}}  & = & 
\frac{1}{l_\delta} \, (I_0[2(k\xi)^2]-I_1[2(k\xi)^2]) \,  
e^{-2 (k\xi)^2} \\ 
& \simeq &
\frac{1}{l_{\rm qm}} \ \frac{1}{4(k\xi)^2} \ .
\hspace{1cm} \ {\rm for} \qquad  k\xi \ \longrightarrow \infty \, .
\end{eqnarray}
This relation shows that $\l_{\rm T}$ can be considerably larger
than $l_{\rm qm}$ for $\lf < \xi$.  This shows that in the case
of a confined system and smooth disorder,
the system may behave ballistically although the elastic MFP $l$ might
be considerably smaller than the system size.


\begin{figure}
\caption{Magnetic susceptibility $\protect \langle \chi \rangle $ 
         (normalized with respect to the Landau susceptibility $
         \chi_L$ of a square billiard as a function of 
         $\protect k_{\scriptstyle F} a$ for the clean 
         case (dotted) and for the ensemble average of billiards of fixed
         size with increasing Gaussian disorder 
         ($\protect \xi/a = 0.1$) according to an elastic mean free 
         path $\protect l/a = 4, 2, 1, 0.5$ (solid lines in the 
         order of decreasing amplitude). The susceptibility is
         calculated for zero magnetic field 
         and at a temperature equal to 6 level spacings.}
\label{fig:chi1}
\end{figure}

\begin{figure}
\caption{Logarithm of the ratio $\protect \langle \chi \rangle 
        / \chi_{cl}$ as a function of the inverse elastic MFP $a/l$. 
         The symbols indicate the numerical 
         quantum results (from the top for $\protect
         \xi/a = 4, 2, 1, 0, 0.5$ and 0.2. The dotted lines show the
         semiclassical analytical results for $\xi/a=4,2,1$ (from
         above) according to Eq.~(\protect\ref{eq:dS2lr}).
         The full line is the semiclassical
         result for $\xi=0$ (Eq.~(\protect\ref{eq:avegrebulk})). 
         The quantum results for $\protect \xi=0.5$ (squares) and
         0.2 (diamonds) are beyond the regime of validity of the 
         analytical limits $\protect \xi/a \gg 1$ and $\xi/a \ll 1$.}
\label{fig:chi1_xidep}
\end{figure}

\begin{figure}
\caption{Averaged magnetic susceptibility (at $H\approx 0$ of an ensemble 
         of square billiards with variations in the size and
         impurity potential ($\xi=0)$ for different 
         disorder strength, i.e.\ elastic mean free path $l_\delta$.
         The full curves show the numerical quantum results
         and the dotted lines the semiclassical predictions
         from Eq.~(\protect\ref{eq:xi0a_dis}) taking into account 
         the variations of $l_\delta$ with $\protect \kf$ (see 
         Eq.~(\protect \ref{eq:sigma_d})). The two sets of curves 
         correspond to an elastic MFP $\protect l_\delta/a=\infty, 8, 4, 2, 1$
         (at $\protect \kf a = 65$), (from the top).}
\label{fig:chia_xi0}
\end{figure}

\begin{figure}
\caption{Logarithm of the ratio between disorder averaged 
         and clean results for (a) typical $\chi^{(\rm t)}$ 
         (b) ensemble averaged $\langle \overline{\chi} \rangle $
         susceptibilities as a function of increasing inverse elastic 
         MFP $a/l$ for different values of $\xi/a$. The symbols denote the 
         numerical quantum results, the solid lines
         (for $ \xi > 0$) the semiclassical integrals 
         (\protect\ref{eq:chit_dis}) (a) and (\protect\ref{eq:chia_dis}) 
         (b) and the dashed lines asymptotic expansions 
         (\protect \ref{eq:db}) 
         of the integrals for large $a/l$.}
\label{fig:chiav}
\end{figure}

\begin{figure}
\caption{ Typical susceptibility as predicted by 
	Eq.~(\protect\ref{eq:chit_dis}) as a function of the dimensionless 
	flux $\varphi=Ha^2/\Phi_0$. Dash line: clean case; solid line:
	$l=a$ and $\xi = 0.1$.}
\label{fig:Hdepend}
\end{figure}

\begin{table}
\begin{tabular}{c|ccc}
         & short-range &  finite-range & long-range \\
\hline 
$\langle\chi\rangle/\chi_{cl}$  & $\exp{(-L_{11}/2l_{\delta})}$   
			     & $\exp{(-L_{11}/2l)}$  
	     & $\exp\left\{-d_1(L^2/l\xi)[1-I_t/(2\xi^2)]\right\}$  \\
$(\chi^{(t)}/\chi^{(t)}_{cl})^2$ & $\exp{(-L_{11}/l_{\delta})}$   
			     & $ c_t (\xi/a)(l/L_{11})^{1/2}$  
			     & $1-d_2 a / l (a/\xi)^9$ \\
$\langle\bar{\chi}\rangle/\bar{\chi}$ & $\exp{(-L_{11}/l_{\delta})}$
				      &$c_a (\xi/a)(l/L_{11})^{1/2}$
				      & $1-d_2 a / l (a/\xi)^9$
\end{tabular}
\caption{Summary of the different average susceptibilities 
(at $\protect H=0$)
 considered in the short-range ($\protect \xi\!<\!\lambda_F\!<\!a$), 
finite-range ($\protect \lambda_F\!<\!\xi\!<\!a$) and long-range ($
\protect \lambda_F\!<\!a\!<\!\xi$) regimes.
The fixed--size impurity averaged susceptibility 
$\protect \langle\chi\rangle$ is given by the one-particle Green function,
while the typical $\protect \chi^{(t)}$ 
and average $\protect \langle\bar{\chi}\rangle$
susceptibilities are given by two-particle Green functions and
involve impurity and energy averages. The different average
susceptibilities are normalized with respect to the corresponding
clean counterparts. $\protect L_{11}$ is the length of the shortest
flux-enclosing periodic trajectories in the square. In the 
short-range regime the damping is governed by the elastic 
mean-free-path $\protect l_{\delta}$ given by the quantum mechanical
expression ({\protect \ref{eq:sigma_d}}).
 The damping in the finite and long--range 
regimes is governed by the elastic MFP $l$ (whose semiclassical 
expression is given in Eq.~(\protect\ref{eq:mfp_gauss})), 
the correlation length $\protect\xi$ of the 
impurity potential and the size 
$a$ of the structure. $\protect I_t$ is the moment of inertia of the 
(11) trajectories (Eq.~(\protect\ref{eq:dS2lr})). The finite--range 
expressions for $\protect\chi^{(t)}$ and 
$\protect \langle \overline{\chi} \rangle$
showing a power--law damping hold in the deep ballistic limit
$\protect l < L_{11}$. The numerical factors are 
$\protect c_t=(20/7)\protect\sqrt{2\pi}, c_a=2\protect\sqrt{2\pi}, 
d_1=1/4\protect \sqrt{\pi}$,  and $\protect d_2 = 6.5 \cdot 10^{-5}$.}
\label{tab:I} 
\end{table}


\begin{references}
\bibitem{foot6}
A proper treatment of these non--diagonal terms in the clean 
limit is an outstanding problem in the context of semiclassical 
quantum transport.
\end{references}
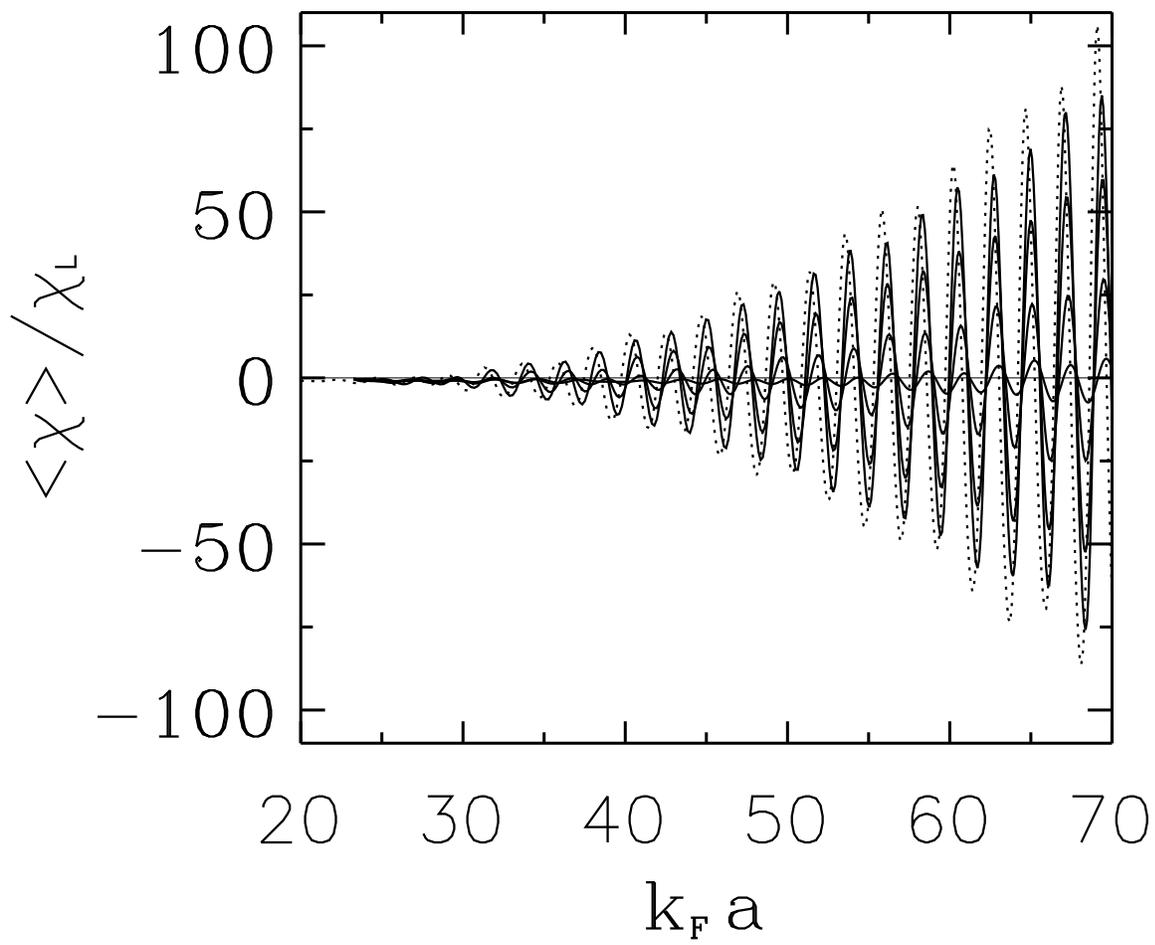
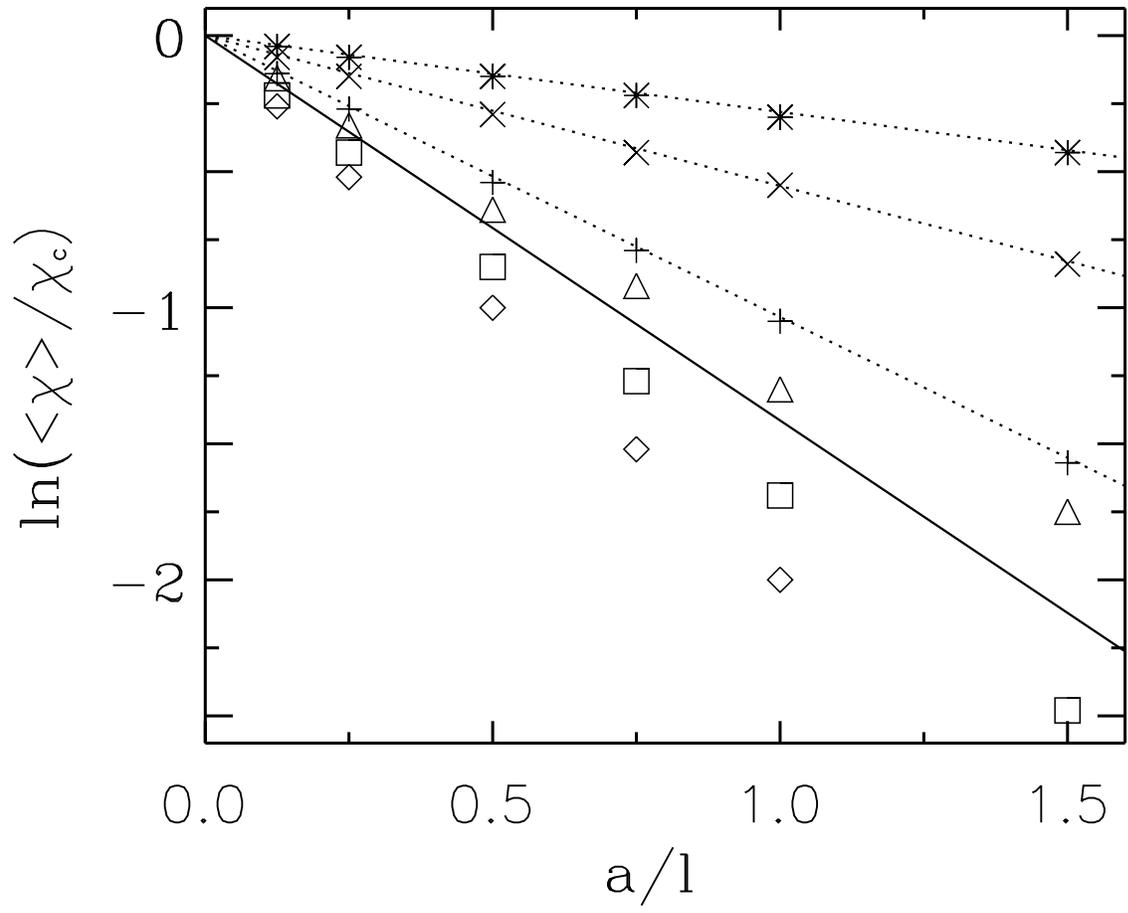
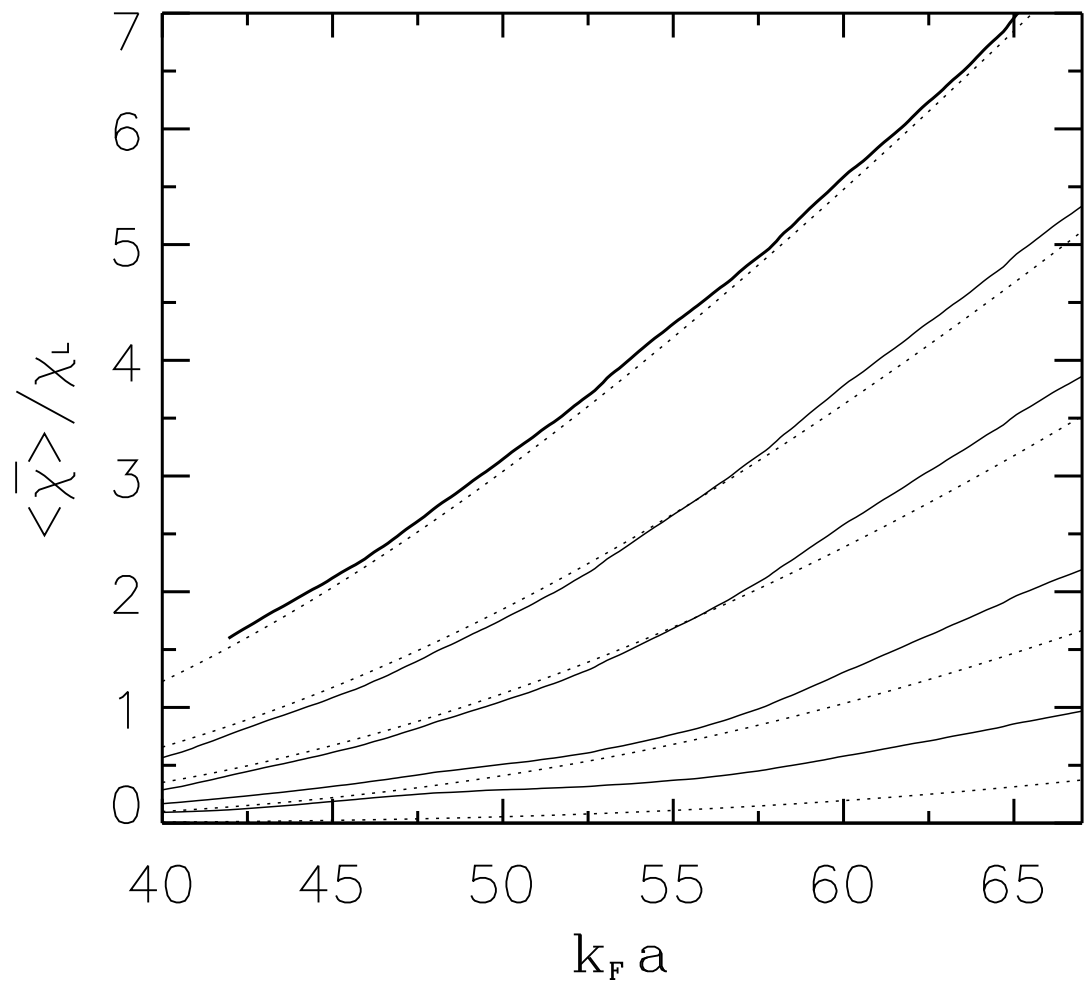
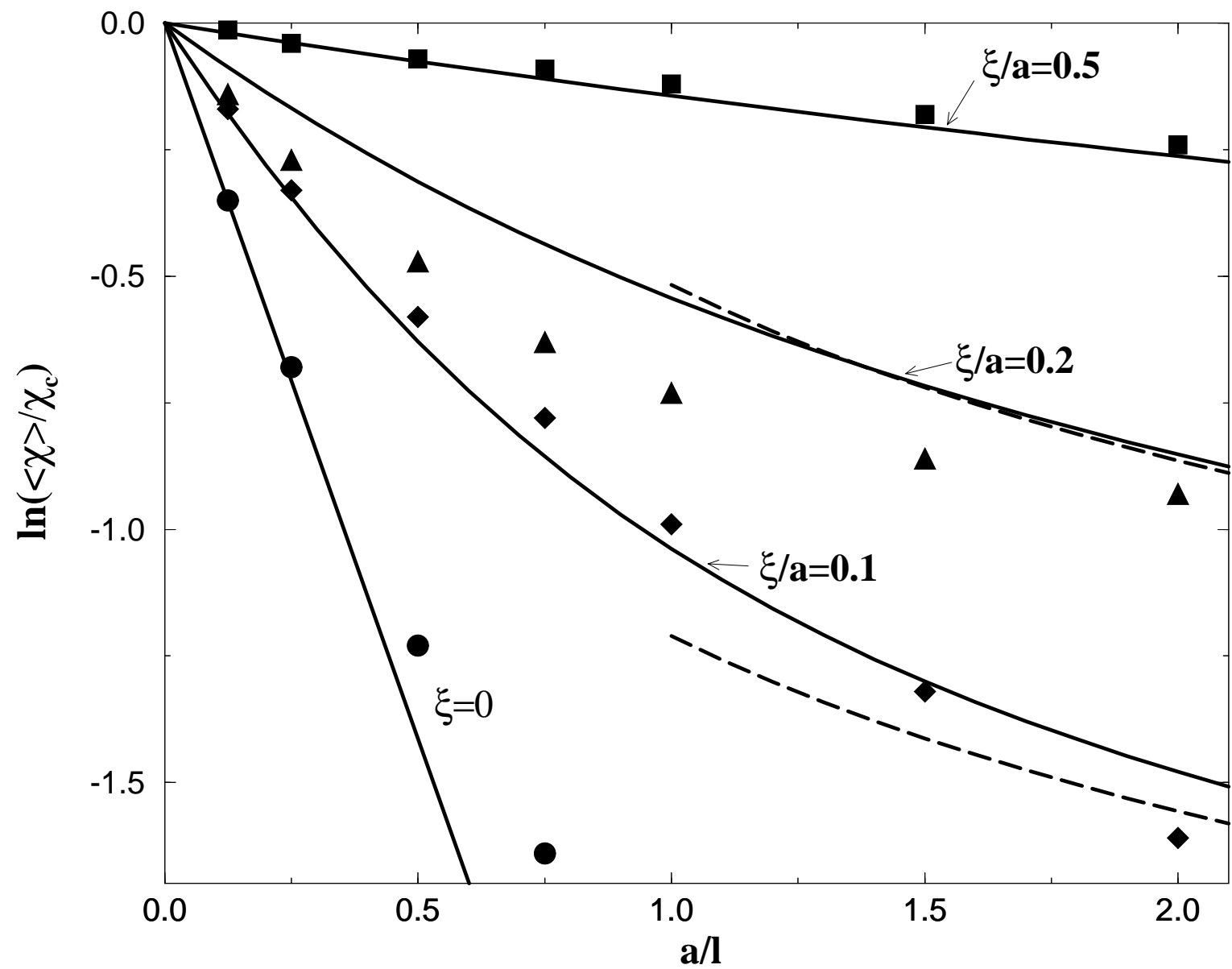
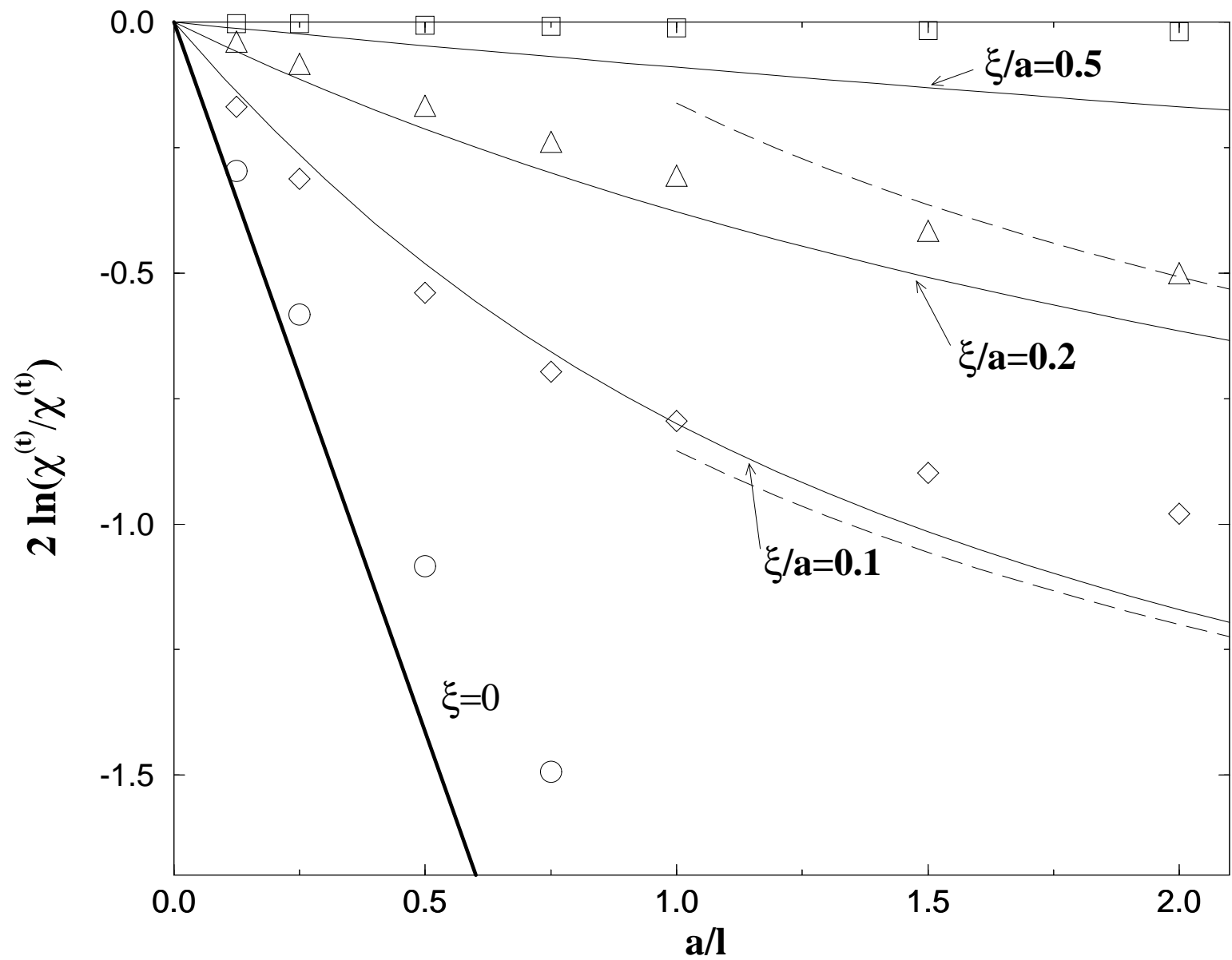
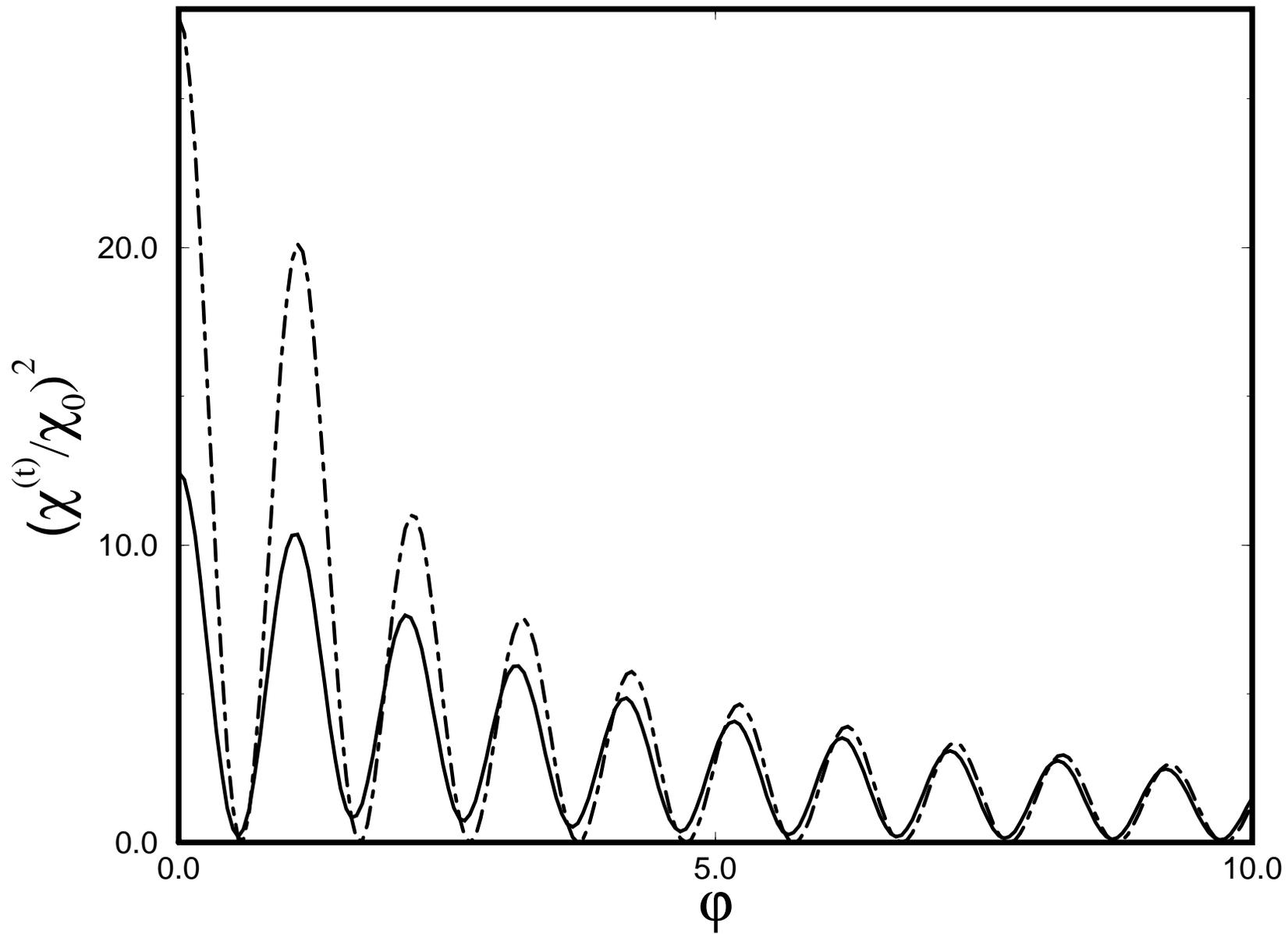
\end{document}